\documentstyle[preprint,eqsecnum,prd,aps,epsf]{revtex}
\tightenlines
\begin{document} 
\input{epsf}
\preprint{
\vbox{
\halign{&##\hfil\cr
	& ANL-HEP-PR-98-92 \cr
	& JLAB-THY-98-43 \cr
	& \today \cr}}}

\title{ Spin Structure of the Proton and Large $p_T$ Processes 
in Polarized $pp$ Collisions}

\author{
L. E. Gordon$^{a,b}$\footnote{
        E-mail address: lgordon@cebaf.gov}
        and G. P. Ramsey $^{c,d}$\footnote{
	Work supported in part by the U.S. Department of Energy,
	Division of High Energy Physics, Contract W-31-109-ENG-38.  
        E-mail address: gpr@hep.anl.gov} }
\address{ $^{a}$  Theory Group, Jefferson Lab, Newport News, VA 23606 \\
$^b$ Hampton University, Hampton, VA 23606 \\
$^{c}$ High Energy Physics Division, Argonne National Laboratory,
	Argonne, IL 60439, USA\\ $^c$ Loyola University, Chicago, IL 60626}
\maketitle
\begin{abstract} 
QCD motivated polarized parton distributions, evolved directly in $x$-space,
are used to predict rates for prompt photon and Jet production at RHIC and 
HERA-$\vec{N}$ center of mass energies. Various scenarios for the polarized 
gluon distributions are considered and compared, and the possibility of 
using large $p_T$ processes in polarized $pp$ collision experiments to 
choose between them is analyzed.  
\end{abstract}
\vspace{0.2in}

\def\ie{{\it i.e.}}
\def\GeV{\,{\rm GeV}}
\def\y{\,{\rm y}}
\def\l{\langle}
\def\r{\rangle}

\setcounter{footnote}{0}
\pagebreak

\section{Introduction}

Interest in spin physics has grown in the last few years thanks to the various
polarized DIS experiments and to existing proposals for future polarized beam
experiments. One proposal includes the construction of a polarized $ep$
collider at DESY (HERA-$\vec{N}$) running at an energy of $\approx 40$ GeV,
which will complement and extend the results to be obtained at the Relativistic
Heavy Ion Collider (RHIC) at Brookhaven.  
Polarized deep-inelastic scattering experiments have shed light on the relative
contributions of the nucleon constituents to their overall spin. However, the
largest theoretical uncertainties are from: (1) the polarized gluon size and
shape as a function of $x$ being unknown and (2) the relative size of the 
polarized sea distributions being uncertain. \cite{gr}

In a recent paper, \cite{ggr} we proposed three new parametrizations for
the spin dependent parton distributions for the proton. Our distributions,
which are available in both leading order (LO) and next-to-leading order (NLO),
were evolved directly in $x$-space. In ref.\cite{ggr} we presented details of
the models used to obtain the input distributions and compared the evolved
structure functions with the available data. All currently available data on
the nucleon spin structure come from deep inelastic scattering experiments,
and therefore do not contain direct information on individual parton
distributions such as the gluon, $\Delta G(x,Q^2)$ or strange sea, 
$\Delta s(x,Q^2)$, for example. It is expected that such detailed information
will come from other experiments such as those proposed at RHIC and
HERA-$\vec{N}$.   

As a $pp$ collider, RHIC will be endowed with a very high integrated luminosity 
(up to 800 pb$^{-1}$ at 500 GeV) and will be spanning a center of mass energy
range between $50$ to $500$ GeV. One of the main programs at RHIC will be to
determine the size and shape of the polarized parton distributions which,
at the moment, suffer from significant model dependence. There are large
uncertainties in both the gluon contribution, $\Delta G$, and the sea quark
distributions.

In this paper we continue the work of ref. \cite{ggr} by providing details of
the $x$-space evolution of the parton distributions. We also compare our
evolved distributions with another parametrization in order to highlight the
similarities and differences between them. Finally, we use our evolved parton
distributions to predict cross sections and asymmetries for direct photon and
single jet production at RHIC and HERA-$\vec{N}$ energies.

\section{Input Distributions}

In reference \cite{ggr}, we constructed three sets of polarized parton
distributions for the valence and all light sea flavors of quarks. Each set
corresponds to a particular model of the polarized gluons, motivated by
different physical assumptions. All of the $x$-dependent polarized quark 
distributions are generated from the unpolarized distributions \cite{CTEQ} and
some basic theoretical assumptions. The valence up and down distributions are
generated from a broken SU(6) model of the polarized valence quarks. For the 
sea distributions, we built in positivity constraints and incorporated the
integrated and $x$-dependent polarized deep-inelastic scattering data. The 
three models of the polarized gluons used to generate the distributions are
assumed small at the relatively low $Q^2$ values of the data. One model gives
a positively polarized gluon, generated from the unpolarized glue: ($\Delta G
= xG$), one is an unpolarized gluon ($\Delta G = 0$), and the third is based
upon an instanton model, which integrates to a slightly negative value. All
three sets of distributions are in good agreement with the data, an indication
that the polarized distributions are still not well determined.

All of our distributions are generated at $Q_0^2=1.0$ GeV$^2$ and are evolved up
to the $Q^2$ values necessary for predicting the appropriate spin observables.
Evolution for the singlet and non-singlet functions is done in $x$-space to
the NLO level. We have assumed three quark flavors up to the appropriate $Q^2$
value for charm production, whereby we smoothly introduce the fourth flavor to
the evolution equations. Details will be discussed briefly in the next section.

The distributions we use here are strongly embedded in sound theoretical
assumptions, while giving good agreement with all of the latest polarized
DIS data. In order to determine the nature of the polarized distributions
to a finer degree of detail, we are using them to generate predictions of
the spin asymmetries for direct photon and jet production. These processes
are feasible for the RHIC and HERA experimental groups and there is general
agreement that they are pivotal in helping to determine the nature of the
polarized gluon distribution.

\section{$x$-Space Evolution in Leading and Next-to-Leading Order}

Given the polarized parton distributions at some reference scale $Q_0^2$, they 
can be predicted at a higher scale $Q^2$ by solving the polarized DGLAP evolution 
equations. For the non-singlet sector the evolution equation is given by
\begin{equation}
\frac{d\Delta q_{NS}(x,Q^2)}{d \ln Q^2}=\frac{\alpha_s(Q^2)}{2\pi}
\Delta P^{\pm}_{NS}(x)\ast \Delta q_{NS}(x,Q^2),
\end{equation}
where $\Delta q_{NS}(x,Q^2)$ is the polarized non-singlet parton 
distribution function, $\Delta P_{NS}^{\pm}(x)$ are polarized non-singlet 
splitting functions, $\alpha_s(Q^2)$ is the strong coupling constant and 
$\ast$ represents the convolution
\begin{equation}
f(x)\ast g(x) =\int^1_x \frac{dy}{y}f\left( \frac{x}{y}\right) g(y).
\end{equation}
For the polarized singlet sector, with the quark singlet defined by
\begin{equation}
\Delta \Sigma =\sum_i (\Delta q_i+\Delta \bar{q}_i), \label{singl}
\end{equation}
the evolution equations are
\begin{eqnarray}
\frac{d\Delta \Sigma(x,Q^2)}{d\ln Q^2} &=&
\frac{\alpha_s(Q^2)}{2\pi}\left[\Delta P_{qq}(x)\ast \Delta \Sigma(x,Q^2) +
\Delta P_{qg}(x)\ast \Delta G(x,Q^2)\right] \nonumber \\
\frac{d\Delta G(x,Q^2)}{d\ln Q^2} &=&
\frac{\alpha_s(Q^2)}{2\pi}\left[\Delta P_{gq}(x)\ast \Delta \Sigma(x,Q^2) +
\Delta P_{gg}(x)\ast \Delta G(x,Q^2)\right].  \label{sevol}
\end{eqnarray}

The splitting functions have the perturbative expansion
\begin{equation}
\Delta P_{ij}(x)=\Delta P^{(0)}_{ij}(x)+\frac{\alpha_s(Q^2)}{2\pi}\Delta 
P^{(1)}_{ij}(x)+...
\end{equation}
The first term is the leading order (LO) contribution and the second term the 
next-to-leading order (NLO) contribution to the splitting kernels, which were 
calculated in ref.\cite{werner}. In leading order we use the expression 
\begin{equation}
\alpha_s(Q^2)=\frac{4\pi}{\beta_0\ln(Q^2/\Lambda^2)},
\end{equation}
and in NLO
\begin{equation}
\alpha_s(Q^2)=\frac{4\pi}{\beta_0\ln(Q^2/\Lambda^2)}\left[1-
\frac{\beta_1\ln{\ln(Q^2/\Lambda^2)}}{\beta^2_0\ln(Q^2/\Lambda^2)}\right]
\end{equation}
for the strong coupling constant.
The constants $\beta_0$ and $\beta_1$ are defined by
\begin{equation}
\beta_0=\frac{11}{3}N_c-\frac{2}{3} N_f, \;\;\;\;\;\ \beta_1=
\frac{34}{3}N_c^2-\frac{10}{3}N_c N_f-2 C_F N_f,
\end{equation}
with the color constants $N_c=3$ and $C_F=4/3$, and $N_f$ is the number of 
active quark flavors. 

In order to solve the DGLAP evolution equations directly in $x$-space we 
use the ansatz proposed in ref.\cite{rossi}. We assume a solution of the 
form
\begin{equation}
\Delta q(x,Q^2)=\sum_{n=0}^{\infty} \frac{A_n(x)}{n!}\ln^n\left( 
\frac{\alpha_s(Q^2)}{\alpha_s(Q^2_0)}\right)+\alpha_s(Q^2)
\sum_{n=0}^{\infty} \frac{B_n(x)}{n!}\ln^n\left( 
\frac{\alpha_s(Q^2)}{\alpha_s(Q^2_0)}\right), \label{esoln}
\end{equation}
where $\Delta q=(\Delta q_{NS},\Delta \Sigma,\Delta G)$. Since the sum 
starts at $n=0$, boundary conditions, $q(x,Q^2_0)=q_0(x)$, can be 
incorporated into the equations. When Eq. (\ref{esoln}) is inserted into the
DGLAP equations and all third-order terms are dropped, the relations
\begin{eqnarray}
A_0(x) & = & q_0(x) \nonumber \\
B_0(x) & = & 0      \nonumber \\
A_{n+1}(x) & = & -\frac{2}{\beta_0} P^{(0)}(x)\ast A_n(x) \nonumber \\
B_{n+1}(x) & = &(-\delta(1-x) -\frac{2}{\beta_0} P^{(0)}(x))\ast B_n(x)-
\frac{1}{\pi\beta_0}P^{(1)}(x)\ast A_n(x)-
\frac{\beta_1}{4\pi\beta_0}A_{n+1}(x)
\end{eqnarray}
are obtained for the $A_n(x)$ and $B_n(x)$. The functions $P^{(0,1)}(x)$ are 
the LO and NLO parts of the splitting functions, while the terms $A_n(x)$ and
$B_n(x)$ represent the LO and NLO coefficients respectively. If a LO evolution 
is desired then only the $A_n(x)$ are calculated and the LO expression for 
$\alpha_s$ is used. 
These equations can be solved numerically for any value of $x$, and the 
coefficients substituted into Eq. (\ref{esoln}) to obtain the distributions
$\Delta q(x,Q^2)$ at any value of $Q^2$. Details of the input distributions
and expressions for the coefficients are given in the Appendix.

\section{Large $p_T$ Processes at RHIC and HERA-$\vec{\rm N}$}

\subsection{Experimental Parameters}

The polarization experiments planned for RHIC show great promise for the high
energy community. With polarized proton beams of 70$\%$ polarization and a
planned luminosity of ${\cal{L}}=2\times 10^{32}$/cm$^{-2}$/sec$^{-1}$, both
direct-$\gamma$ production and jet production can be done in a kinematic region
where we can gain tremendous insight into the polarized gluon distribution.
These are part of the planned set of experiments for both the STAR and PHENIX
detectors at RHIC. \cite{bnl} If the experiments are able to obtain an
integrated luminosity, presently assumed to be 320 $pb^{-1}$ at $\sqrt{s}=200
\GeV$ and 800 $pb^{-1}$ at $\sqrt{s}=500 \GeV$, the resulting data should be
enough to distinguish between various polarized gluon models which have been
proposed. \cite{ggr,gluon}

The STAR detector will have a wide angular coverage to cover a large kinematic
region for Jet production. Their range of $x$ will extend from
about $0.02\leq x\leq 0.30\> (\pm 0.02)$ with an approximate uncertainty of
$\delta \l \Delta G\r \approx \pm 0.05$.

The PHENIX detector will measure high $p_T$ prompt $\gamma$'s in the kinematic
range of $10 \leq p_T\leq 30 \GeV$ at $\sqrt{s}=200 \GeV$ ($x\leq 0.25$) and
$10 \leq p_T\leq 40 \GeV$ at $\sqrt{s}=500 \GeV$ ($x\leq 0.15$). The angular
coverage is not as large as STAR but it has finer granularity. The estimated
uncertainty is $\delta \l {{\Delta G}\over G} \r \approx \pm 0.05\to
0.30$, increasing linearly with $p_T$. The combination of the two detectors
will be valuable in covering a wide kinematic range with reasonable error
bars for cross sections and asymmetries.

Tests are presently being performed to polarize the proton beam at HERA.
The success of these tests will provide a unique kinematic region for measuring
asymmetries in direct $\gamma$ and Jet production. \cite{nowak}
These experiments will be performed at $\sqrt{s}=40 \GeV$, giving a good
sensitivity to distinguishing between gluon model predictions for the
corresponding asymmetries. This will be discussed later. 

All of the above mentioned experimental groups plan to measure the observables
in both direct-$\gamma$ and Jet production. Present plans call for
the RHIC experiments to be performed in the next five years, while the HERA
experiments would hopefully follow shortly thereafter. Both sets of experiments
are important, as the lower center-of-mass energies appear to be somewhat
better for distinguishing the size of the polarized gluons. Thus, the
combination of a wide kinematic range and reasonable uncertainties should
give a good indication of the size and shape of the polarized gluon distribution.

\subsection{Theoretical Background}

\subsubsection{Prompt Photon Production}

The prompt photon cross section promises to be one of the most useful for
measuring $\Delta G$ at RHIC and HERA-$\vec{N}$ since it is dominated by
initial $qg$ scattering. Contributions to the prompt photon cross section
are usually separated into two classes in both LO \cite{bergerqiu} and NLO.
\cite{contogouris,gorvogel,gordon1}
There are the so-called direct processes, $ab\rightarrow \gamma c$ in LO
and $ab\rightarrow \gamma cd$, in NLO,
$a,b,c$ and $d$ referring to partons, where the photon is produced directly
in the hard scattering. In addition there are the fragmentation contributions
where the photon is produced via bremsstrahlung off a final state quark or
gluon, $ab\rightarrow cd(e)$ followed by $c\to \gamma + X$ for instance. 

In LO, $O(\alpha \alpha_s)$, the direct subprocesses contributing 
to the cross section are
\begin{eqnarray}
qg&\rightarrow& \gamma q \nonumber \\
q\bar{q}&\rightarrow& \gamma g. \label{bproc}
\end{eqnarray}
In addition there are the fragmentation processes
\begin{eqnarray}
qg &\rightarrow& q g \nonumber \\
qq &\rightarrow& q q \nonumber \\
qq' &\rightarrow& q q' \nonumber \\
q\bar{q} &\rightarrow& q \bar{q} \nonumber \\
qg &\rightarrow& q g \nonumber \\
q\bar{q} &\rightarrow& g g \nonumber \\
gg &\rightarrow& g g \nonumber \\
gg &\rightarrow& q \bar{q} \label{fproc}
\end{eqnarray}
where one of the final state partons fragments to produce the photon,
\ie, $q (g)\rightarrow \gamma + X$.

In the direct processes in LO, the differential cross section is given by
\begin{equation}
E_\gamma\frac{d\Delta\sigma_{dir}^{LO}}{d^3p_\gamma}=\frac{1}{\pi S}\sum_{i,j}
\int^V_{V
W}\frac{dv}{1-v}\int^1_{VW/v}\Delta f^i_1(x_1,M^2)\Delta f^j_2(x_2,M^2)
\frac{1}{v}\frac
{d\Delta\hat{\sigma}_{ij\rightarrow\gamma}}{dv}\delta(1-w)
\end{equation}
where $S=(P_1+P_2)^2$, $V=1+T/S$, $W=-U/(T+S)$, $v=1+\hat{t}/\hat{s}$,
$w=-\hat{u}/(\hat{t}+\hat{s})$, $\hat{s}=x_1 x_2 S$, and
$T=(P_1-P_\gamma)^2$ and $U=(P_2-P_\gamma)^2$. As usual the Mandelstam
variables are defined in the upper case for the hadron-hadron system
and in lower case in the parton-parton system. $P_1$ and $P_2$ are the 
momenta of the incoming hadrons and $\Delta f^i_1(x_1,M^2)$ and
$\Delta f^j_2(x_2,M^2)$ represent the respective probabilities of finding
parton $i$ and $j$ in hadrons $1$ and $2$ with momentum fractions $x_1$ and
$x_2$ at scale $M^2$. The quantity $d\Delta \hat{\sigma}_{ij\to \gamma}/dv$
is the hard subprocess cross section which is calculable in perturbative QCD.
 
For the fragmentation processes, the differential cross section is given by
\begin{eqnarray}
E_\gamma\frac{d\Delta\sigma_{frag}^{incl}}{d^3p_\gamma}&=&\frac{1}{\pi S}
\sum_{i,j,l}\int^1_{1-V+VW}\frac{dz}{z^2}\int^{1-(1-V)/z}_{V
W/z}\frac{dv}{1-v}\int^1_{VW/vz}\frac{dw}{w}\Delta f^i_1(x_1,M^2)\Delta 
f^j_2(x_2,M^2) \nonumber \\
&\times &\frac{1}{v}
\frac{d\Delta\hat{\sigma}_{ij\rightarrow l}}{dv}\delta(1-w)D_{\gamma/l}
(z,M_f^2),
\end{eqnarray}
where $D_{\gamma/l}(z,M_f^2)$ represents the probability that the parton
labelled $l$ fragments to a photon with a momentum fraction $z$ of its
own momentum at scale $M_f^2$ (note that $D_{\gamma/l}(z,M_f^2)$ is the
usual unpolarized fragmentation function, since the final state is not
polarized). This is the non-perturbative fragmentation function which must
be obtained from experiment at some scale and evolved to $M_f^2$ using the
usual evolution equations.  

In NLO, $O(\alpha\alpha_s^2)$, there are virtual corrections to the LO
non-fragmentation processes of Eq. (\ref{bproc}), as well as the further
three-body processes:
\begin{mathletters}\label{eq:1}
\begin{eqnarray}
g &+q  \rightarrow g + q + \gamma\label{eq:11}\\
g &+ g \rightarrow q +\bar{q} + \gamma\label{eq:12}\\
q &+ \bar{q} \rightarrow g + g + \gamma\label{eq:13}\\
q &+ q \rightarrow  q + q + \gamma\label{eq:14}\\
\bar{q} &+ q \rightarrow \bar{q} + q + \gamma\label{eq:15}\\
q &+ \bar{q} \rightarrow q' + \bar{q}' + \gamma\label{eq:16}\\
q &+ q' \rightarrow q + q' + \gamma. \label{eq:17}
\end{eqnarray}
\end{mathletters} 

In principle the the fragmentation processes of Eq. (\ref{fproc}) should now be
calculated to $O(\alpha_s^3)$ and convoluted with the NLO photon
fragmentation functions whose leading behaviour is $O(\alpha/\alpha_s)$,
but they have not yet been calculated for the
polarized case, hence, in both the polarized and unpolarized cases, we
include the leading order contributions to these processes only.

The direct contribution to the inclusive cross section is given by
\begin{eqnarray}
E_\gamma\frac{d\Delta\sigma^{incl}_{dir}}{d^3p_\gamma}&=&\frac{1}{\pi S}
\sum_{i,j}\int^V_{V
W}\frac{dv}{1-v}\int^1_{VW/v}\frac{dw}{w}\Delta f^i_1(x_1,M^2)\Delta 
f^j_2(x_2,M^2)
\nonumber \\ &\times&\left[ \frac{1}{v}\frac
{d\Delta\hat{\sigma}_{ij\rightarrow\gamma}}{dv}\delta(1-w)+
\frac{\alpha_s(\mu^2)}{2\pi}
\Delta K_{ij\rightarrow\gamma}(\hat{s},v,w,\mu^2,M^2,M_f^2)\right],
\end{eqnarray}
where $\Delta K_{ij\rightarrow\gamma}(\hat{s},v,w,\mu^2,M^2,M_f^2)$ represents
the higher corrections to the hard subprocess cross sections calculated
in \cite{gorvogel} and $\mu$ is the renormalization scale. 

In this paper we present results for the inclusive prompt photon cross section
without taking any possible isolation cuts into account. As shown in
ref.\cite{gordon1}, isolation cuts do not have a significant effect on
the asymmetries, which are the quantities in which we are mainly
interested here. It is also unlikely that isolation will be necessary at
HERA cms energies.
 
\subsubsection{Single Jet Production}

 We calculate the single jet cross section at LO only. In this case the
approximation `parton = jet' is used and a jet definition is not needed.
The eight subprocesses listed in Eq. (\ref{fproc}) are the same ones
contributing to jet production in LO. The differential cross section is
given by;
\begin{equation} 
E_J\frac{d\Delta\sigma^{J}}{d^3p_J}=\frac{1}{\pi S}\sum_{i,j}
\int^V_{V
W}\frac{dv}{1-v}\int^1_{VW/v}\Delta f^i_1(x_1,M^2)\Delta f^j_2(x_2,M^2)
\frac{1}{v}\frac
{d\Delta\hat{\sigma}_{ij\rightarrow J}}{dv}\delta(1-w)
\end{equation} 
where the variables are defined as before. \cite{gordon2} Recently the
polarized jet cross section has been calculated in NLO, \cite{nlojet} but 
for the purposes of this paper we use the leading order calculation.

\subsection{Significance of Results}

We now compare our three sets of polarized distributions, (GGRA, GGRB and GGRC)
\CITE{ggr} generated from the three gluon models discussed in section II, with
those of Gehrmann and Stirling,\cite{gs} set GSA. Our polarized gluon
distributions are smaller than those of GSA, which affects the ratio of quark
to gluon contributions at small $x$. This in turn will modify the relative
contributions to direct-$\gamma$ production and Jet production. (See
fig. 1a). The polarized sea peaks at higher $x$ and goes negative at much larger
$x$ than the GSA distributions, which will further affect the relative
contributions at small $x$. (See figs. 1b and 1c). Our valence distributions
are also larger, which makes the qq contributions to both processes larger.
(See fig. 1d). For GGRA, the LO and NLO distributions are similar, implying 
perturbative stability. The result is similar for GGRB and GGRC (figs. 1c and
1d).  

The comparative results for direct-$\gamma$ production are shown in figures
2 through 4. At $\sqrt{S}=200 \GeV$, the cross section is measurable out to
$p_T\sim 40-50\GeV$, given the planned luminosities of RHIC (fig. 2a). The
$A_{LL}^{\gamma}$ asymmetry in direct-$\gamma$ production differs significantly
from GSA only at the largest $p_T$ values ($>40 \GeV$), where the PHENIX
detector at RHIC is at an upper $p_T$ limit. However, expected errors in the
measurement of the asymmetry, $A_{LL}^{\gamma}$, at PHENIX are much smaller at
the lower $p_T$ values. \cite{goto} Up to $p_T$ of about $15\GeV$, the expected
error is about $\pm 0.006$ and grows to about $\pm 0.080$ at $30\GeV$. Thus,
measurement of this asymmetry should be done at all possible $p_T$ values,
with those at lower $p_T$ possibly able to distinguish between the small and
large polarized gluon distributions, despite their close proximity. (See fig.
2b). At $\sqrt {S}=200\GeV$, the $qg$ processes in GGRA dominate (positive
$\Delta G$) much more than the GSA and our other gluon models, which give the
larger $A_{LL}^{\gamma}$ values seen at larger $p_T$. The $q\bar{q}$
contributions are relatively small at all $p_T$ for all models. See fig. 2c
for details. Thus, the $qg$ processes dominate, as expected. In fig. 3a, we
plot the direct-$\gamma$ cross section as a function of rapidity,
$y^{\gamma}$ at $p_T^{\gamma}=15\GeV$. The cross section remains of reasonable
size out to rapidities past $\mid y^{\gamma}\mid \approx 1.3$. The 
corresponding asymmetries (fig. 3b) are relatively flat out to
$\mid y^{\gamma}\mid \approx 1$, after which they exhibit a slight rise. They
become quite large at rapidities which are out of the range for PHENIX, but
possibly accessible by the endcap of the STAR detector. It may be advantageous
to take measurements at $\mid y^{\gamma}\mid \geq 1$ at the appropriate ranges
of rapidity with both detectors for better sensitivity to the asymmetries.

At HERA energies, $\sqrt {s}\approx 40 \GeV$, the cross section is reasonable
for ranges of $p_T$ up to about $10 \GeV$ (fig. 4a). There are differences in
the asymmetry $A_{LL}^{\gamma}$, which depend upon the models of the polarized
distributions. However, these differences manifest themselves more greatly at
$p_T \approx 10 \GeV$ and above. The sizes of $A_{LL}^{\gamma}$ are still
large enough at the planned $p_T$ values at HERA (about $2\to 8 \GeV$) to be
measured. Since the asymmetries diverge quicker at the HERA center-of-mass
energy than at higher $\sqrt {s}$, there may be a good chance that a more
precise determination of the polarized gluon can be made. A lot depends upon
the relative errors expected from such measurements at each $p_T^{\gamma}$.
(See fig. 4b). Thus, even the moderately sized positive $\Delta G$ gives a 
relatively large asymmetry, $A_{LL}^{\gamma}$. Small asymmetries would
indicate a very small, if not negative $\Delta G$ in this kinematic region.
A possible problem exists, since the expected experimental errors tend to
increase substantially with increasing $p_T$. Thus, these two cases may not
be distinguishable. It is recommended that $A_{LL}^{\gamma}$ be measured over
a wide range of $p_T$, up to the largest values possible, to test the
consistency of these models.

Although the direct-$\gamma$ cross sections are significantly larger at 
the RHIC energy of $\sqrt {S}=500\GeV$, the asymmetries are correspondingly
smaller and it would be much more difficult to determine the polarized gluon
distribution from direct-$\gamma$ production. Thus, we highly recommend that
these measurements be done at HERA-$\vec{N}$ and at the lower $\sqrt {S}$
values at RHIC.

Figures 5 through 6 show our predictions for jet production ($pp\to$
Jet+X) using these distributions. The cross sections are measurable for
all the planned values of $p_T^J$. These are shown for zero rapidity
at $\sqrt {S}=200 \GeV$ in fig. 5a. The $qq$ and $qg$ processes both
dominate in our model for $A_{LL}^J$ at large $p_T^J$. The $qg$ is more dominant
at smaller $p_T^J$, with comparable contributions at the limit of the RHIC
STAR detector, ($p_T\approx 25 \GeV$ at $\sqrt {S}=200\GeV$). This is
consistent with the discussion above. The GSA model has the $qg$ process 
totally dominant with $qq$ small (see fig. 5b). However, their asymmetries
are quite a bit smaller than ours for $p_T^J \ge 20 \GeV$ and ours rise with
$p_T^J$, while theirs remain constant. A measurement of $A_{LL}^J$ for
different $p_T^J$ at the smaller $\sqrt{S}$ region of RHIC may be able to
distinguish among these models, and give an indication as to which process
does indeed dominate in this kinematical region. The GS and GGR models are both
around the $4\to 6\%$ level at $p_T^J\approx 20 \GeV$ while the GGR asymmetries
grow with $p_T^J$ up to the maximum values at RHIC (see fig. 5c). Our
asymmetries for the other gluon models exhibit the same $p_T^J$ behavior,
except that the values are smaller (up to about the $10\%$ level at the limit).
Again, it is important that the errors in jet measurement at STAR be minimized,
in order to distinguish among the models at these levels. This can be, at least
in part, accomplished with the endcap installed, which will provide enhanced
jet coverage in this kinematic region. Distinguishing among the various models
can yield vital information about both the polarized gluons and quark
distributions in Jet production. (See figs. 5d and 5e).

At large $\sqrt{s}\approx 500 \GeV$, the cross sections are larger than at $200
\GeV$ (see fig. 6a), but the corresponding asymmetries in Jet
production are much smaller and cannot distinguish between the various models,
(fig. 6b) even at the largest $p_T^J$ values available. Thus, it is recommended
that these asymmetries be measured at the lower center-of-mass energies.

\section{Discussion}

We have calculated the cross sections and asymmetries for polarized
direct-$\gamma$ production and Jet production at RHIC and HERA (for $\gamma$
production) cms energies, using the polarized distributions generated in
reference \cite{ggr}. Comparisons of our predictions with those of other
distributions were discussed in light of the physics that can be extracted
from performing these experiments. The entire $Q^2$ evolution for these
predictions was done for LO and NLO in $x$-space using a technique explained 
in the text and in the Appendix. This gives highly reliable results without
the use of moments. Both the LO and NLO evolution programs are written in
FORTRAN and are available from either author via electronic mail.

We have shown that if the polarized gluon distribution is even moderately
positive at $Q^2=1\GeV$, the asymmetry for direct photon production is among
the best candidates for determining the size of $\Delta G$. The size of the
predicted asymmetries may, however, be less relevant if the experimental
errors are too large. Jet production is a close contender for determining
which models of $\Delta G$ give the best agreement with data. Much will depend 
upon the relative errors in the applicable kinematic regions for PHENIX (photon
detection) and STAR (detection of photons/jets). For details, see the articles
by Y. Goto, S. Vigdor, S. Heppelman and D. Underwood in ref. \cite{bnl}
At this point, it appears that the projected errors for direct-$\gamma$
production at $\sqrt S=200\GeV$ may be able to distinguish between predictions
based upon large and small values of $\Delta G$. Any sensitivity of polarized
direct-$\gamma$ production to $k_T$ smearing effects should cancel out
when the asymetries are calculated, which
should eliminate some of the potential theoretical uncertainties. Furthermore,
using both PHENIX and STAR to measure direct photon production will enlarge
the rapidity range, and thus increase the sensitivity to extract useful physics
information from this process.

We have also shown that Jet production provides an excellent mechanism to
distinguish among polarized gluon models, as well as polarized quark
distribution models. The corresponding predicted asymmetries are large enough
to measure and distinct enough to distinguish among many models, including
those mentioned in the last section and other extreme models, such as a
saturated $\Delta G$ model. \cite{soffer} It is important to stress that it is
necessary to include both triggering and endcap detection in STAR for the 
fullest coverage of $x$. This will help not only determine the size, but also
the shape of $\Delta G$. The jet production rates appear to be sufficient in
the appropriate kinematic regions,\cite{st} but detection must be complete to
extract the maximum physics information. 
The two processes discussed above seem to have greater size and more
significant differences in the various predicted asymmetries than J/$\psi$
production for the same kinematic regions. \cite{tt} It is thus important
that experimental errors be minimized for both PHENIX and STAR so that these
measurements can be made.

Since HERA-$\vec{N}$ coverage for direct-$\gamma$ production is at lower cms
energy and spans a lower $p_T$ range, this experiment would be complementary
with that of RHIC in the gluon kinematical region covered. The sensitivity of
HERA-$\vec{N}$ to $\Delta G$ for direct-$\gamma$ production is sufficient to
distinguish between the various gluon models. \cite{gv2} Coupling this with
the other planned experiments at HERA-$\vec{N}$ would span a sufficient
kinematic range to give a very good indication as to both the size and shape
of $\Delta G$. \cite{ddhlr}

There are significant enough differences in the models for the gluon
distributions \cite{ggr,gluon,gs,soffer} to warrant careful study of these
processes. Measurement of asymmetries for direct photon production and jet
production at RHIC and HERA-$\vec {N}$ are crucial in determining the size
and shape of the polarized gluon distribution. 

\section*{Acknowledgments}

The work at Argonne National Laboratory was supported by the US Department of
Energy, Division of High Energy Physics, Contract number W-31-109-ENG-38.
LEG acknowledges the hospitality of the FNAL theory group where part of this
work was completed. GPR gratefully acknowledges discussions on experimental
parameters with A. Vasiliev and S. Vigdor.

\begin{appendix}
\section*{x-Space Evolution}
 In this appendix we give details of our $x$-space evolution of the 
polarized parton densities. We use the definitions of ref.\cite{werner} for 
the non-singlet distributions and splitting functions.

Starting at $Q_0$ with $N_f=3$ flavours we define non-singlet distributions 
\begin{equation}
\Delta q_i^{\pm}=\Delta q_i \pm \Delta \bar{q}_i
\end{equation}
and evolve the distributions $\Delta q_i^-$ and $\Delta q_i^+-\Delta q_j^+$ 
according to the equations
\begin{equation}
\frac{d(\Delta q_i^+ -\Delta q_j^+)(x,Q^2)}{d \ln
Q^2}=\frac{\alpha_s(Q^2)}{2\pi}
\Delta P^{+}_{qq}(x)\ast (\Delta q_i^+ -\Delta q_j^+)(x,Q^2)
\end{equation}
and
\begin{equation}
\frac{d\Delta q_i^-(x,Q^2)}{d \ln Q^2}=\frac{\alpha_s(Q^2)}{2\pi}
\Delta P^{-}_{qq}(x)\ast \Delta q_i^-(x,Q^2)
\end{equation}
where
\begin{equation}
\Delta P_{qq}^{\pm}=P^V_{qq}\pm\Delta P^V_{q\bar{q}}.
\end{equation}

The singlet distribution is defined in Eq. (\ref{singl}) and the evolution
equation is given by Eq. (\ref{sevol}). In ref. \cite{werner} the definition
\begin{equation}
\Delta P_{qq}=\Delta P^+_{qq}+\Delta P^S_{qq}
\end{equation}
is used.
All the splitting functions are given in that reference.

For $N_f=3$ flavours we have four non-singlet distributions to evolve
\begin{eqnarray}
T_3 & = & \Delta u^+-\Delta d^+ = \Delta u+\Delta \bar{u}- \Delta d-
\Delta \bar{d}
\nonumber \\
T_8 & = &  \Delta u^++\Delta d^+-2\Delta s^+ = \Delta u+\Delta \bar{u}+
\Delta d+\Delta \bar{d} -4\Delta s \nonumber \\
V_1 & = & \Delta u - \Delta \bar{u} \nonumber \\
V_2 & = & \Delta d - \Delta \bar{d}.
\end{eqnarray}
plus the singlet quark
\begin{equation}
\Sigma = \Delta u+\Delta \bar{u}+\Delta d+\Delta \bar{d} +2\Delta s 
\end{equation}
and gluon, $\Delta G$, distributions. We have assumed that $\Delta s = 
\Delta \bar{s}$. The functional forms of the input distributions are given 
in reference \cite{ggr}.

We label the coefficient functions for the different sectors by
$A_n^{NS,+},A_n^{NS,-},A_N^{\Sigma}$ and $A_n^G$ and correspondingly for the 
$B's$. In LO the coefficients are given by the relations;

\begin{eqnarray}
A_{n+1}^{NS,\pm}(x) & = &A_{n}^{NS,\pm}(x)\left[ -\frac{3C_F}{\beta_0}- 
\frac{4C_F}{\beta_0}\ln (1-x) \right] \nonumber \\
& + & \frac{2C_F}{\beta_0}\int^1_x \frac{dy}{y} (1+z)A_{n}^{NS,\pm}(y)-
\frac{4 C_F}{\beta_0} \int^1_x \frac{dy}{y} 
\frac{y A_{n}^{NS,\pm}(y) - x A_{n}^{NS,\pm}(x)}{y-x} 
\end{eqnarray}

\begin{eqnarray}
A_{n+1}^{\Sigma}(x) & = & A_{n}^{\Sigma}(x)\left[ -\frac{3C_F}{\beta_0}- 
\frac{4C_F}{\beta_0}\ln (1-x) \right] \nonumber \\
& + & \frac{2}{\beta_0}\int^1_x \frac{dy}{y}\left\{ C_F (1+z)A_{n}^{\Sigma}(y)-
N_f(2z-1)A^G_{n}(y) \right\} \nonumber \\
& - & \frac{4 C_F}{\beta_0} \int^1_x \frac{dy}{y} 
\frac{y A_{n}^{\Sigma}(y) - x A_{n}^{\Sigma}(x)}{y-x} 
\end{eqnarray}

\begin{eqnarray}
A_{n+1}^{G}(x) & = &  - A_{n}^{G}(x)\left[1 + \frac{4 N_c}{\beta_0}\ln (1-x) 
\right] \nonumber \\
& - & \frac{2}{\beta_0}\int^1_x \frac{dy}{y}\left\{ (2-z)A_{n}^{\Sigma}(y)-
\frac{4 N_c}{\beta_0} (2 z-1) A^G_n(y) \right\} \nonumber \\
& + & \int^1_x \frac{dy}{y} 
\frac{y A_{n}^{G}(y) - x A_{n}^{G}(x)}{y-x} 
\end{eqnarray}
where $z=x/y$.

The NLO coefficients are given by;

\begin{eqnarray}
 B^{NS,\pm}_{n+1}(x) & = & -\frac{\beta_1}{4\pi\beta_0} A^{NS,\pm}_{n+1}(x)
 -B^{NS,\pm}_n(x)\left[ 1 + \frac{C_F}{\beta_0}(3 + 4
\ln(1-x) \right] + \frac{C_F}{\pi\beta_0}\left[ N_F\left( \frac{1}{12}+
\frac{\pi^2}{9} \right. \right. \nonumber \\
& + & \left. \left. \frac{10}{9}\ln(1-x) \right)
 -  C_F\left( \frac{3}{8}-\frac{\pi^2}{2}+6\zeta \right)-N_c\left( 
\frac{17}{24}+\frac{11\pi^2}{18}-3\zeta+\left( \frac{67}{9}\right.
\right. \right. \nonumber \\
& -& \left. \left. \left.  \frac{\pi^2}{3}
\right) \ln(1-x) \right) \right] A^{NS,\pm}_n(x) 
 +  \int^1_x \frac{dy}{y}\left[ 
-\frac{4 C_F}{\beta_0}\frac{y B^{NS,\pm}_n(y)-x B^{NS,\pm}_n(x)}{y-x}
\right. \nonumber \\ 
&+& \left.  \frac{2 C_F}{\beta_0}(1+z) B^{NS,\pm}_n(y) 
 +  \frac{C_F}{\beta_0\pi} \left( \frac{10}{9}N_f  
 -  N_c\left( 
\frac{67}{9}-\frac{\pi^2}{3} \right) \right) \right. \nonumber \\
& & \left. \frac{y A^{NS,\pm}_n(y)-x A^{NS,\pm}_n(x)}{y-x}
 - \frac{1}{\pi\beta_0}
\Delta P'^{(1),{\pm}}_{qq}(z) A^{NS,\pm}_n(y) 
\right], 
\end{eqnarray}

\begin{eqnarray}
 B^{\Sigma}_{n+1}(x) & = & -\frac{\beta_1}{4\pi\beta_0} A^{\Sigma}_{n+1}(x)
 -B^{\Sigma}_n(x)\left[ 1 + \frac{C_F}{\beta_0}(3 + 4
\ln(1-x) \right] + \frac{C_F}{\pi\beta_0}\left[ N_F\left( \frac{1}{12}+
\frac{\pi^2}{9} \right. \right. \nonumber \\
& + & \left. \left. \frac{10}{9}\ln(1-x) \right)
 -  C_F\left( \frac{3}{8}-\frac{\pi^2}{2}+6\zeta \right)-N_c\left( 
\frac{17}{24}+\frac{11\pi^2}{18}-3\zeta+\left( \frac{67}{9}\right.
\right. \right. \nonumber \\
& -& \left. \left. \left.  \frac{\pi^2}{3}
\right) \ln(1-x) \right) \right] A^{\Sigma}_n(x) 
 +  \int^1_x \frac{dy}{y}\left[ 
-\frac{4 C_F}{\beta_0}\frac{y B^{\Sigma}_n(y)-x B^{\Sigma}_n(x)}{y-x}
\right. \nonumber \\ 
&+& \left.  \frac{2 C_F}{\beta_0}(1+z) B^{\Sigma}_n(y) 
 +  \frac{C_F}{\beta_0\pi} \left( \frac{10}{9}N_f  
 -  N_c\left( 
\frac{67}{9}-\frac{\pi^2}{3} \right) \right) 
\frac{y A^{\Sigma}_n(y)-x A^{\Sigma}_n(x)}{y-x}\right. \nonumber \\
& - & \left. \frac{1}{\pi\beta_0}
\left( \Delta P^{(1)}_{qg}(z) A^{G}_n(y)+\Delta P_{qq}^{+}(z)A^{\Sigma}_n(y)
\right) \right. \nonumber \\
& - & \left. \frac{2 N_f}{\beta_0}(2z-1) B_n^{G}(y)
 - \frac{C_F N_f}{\pi\beta_0}\left\{  (1-z)-(1-3z)\ln z\right. \right.
\nonumber \\
& - & \left. \left. (1+z)\ln^2 z\right\} A^{\Sigma}_n(y) 
\right], 
\end{eqnarray}

\begin{eqnarray}
 B^G_{n+1}(x) &=& -\frac{\beta_1}{4\pi\beta_0} A^G_{n+1}(x)-B^G_n(x)
\left(2 + \frac{4 N_c}{\beta_0}\ln(1-x) \right) - \frac{1}{\pi\beta_0}
A^G_n(x)\left[ N_c^2 \left( \frac{8}{3}\right. \right. \nonumber \\
&+& \left. \left. 3\zeta \right)-\frac{C_F N_f}{2}-
\frac{2 N_c N_f}{3}+\left\{ N_c^2 \left( \frac{67}{9}-\frac{\pi}{3}\right)
\right. \right. \nonumber \\
&- & \left. \left. 
\frac{10 N_f N-}{9}\right\} \ln(1-x) \right]
+\int^1_x \frac{dy}{y}\left[ 
 -\frac{4 N_c}{\beta_0}\frac{y B^G_n(y)- x B^G_n(x)}{y-x} \right.
\nonumber \\
&- & \left. \frac{2 C_F}{\beta_0} (2-z)B^{\Sigma}_n(y) 
+\frac{4 N_c}{\beta_0}(2z-1)B^G_n(y)-\frac{1}{\pi\beta_0}
\left( N_c^2\left( \frac{67}{9}-\frac{\pi^2}{3}\right) 
\right. \right. \nonumber \\
&- &\left. \left. \frac{10 N_c}{9} 
\right)\frac{y A^G_n(y)-x A^G_n(y)}{y-x}-\frac{1}{\pi\beta_0}
(\Delta P^{(1)}_{gq}(z) A^{\Sigma}_n(y)+\Delta P'^{(1)}_{gg}(z)
A^G_n(y))\right],
\end{eqnarray}

where

\begin{eqnarray}
P^+_{qq}(z)&=& C_F^2\left\{ -\left( 2 \ln(z) \ln(1-z)+\frac{3}{2}\ln(z)
\right) \delta P_{qq}(z)-\left(\frac{3}{2}+\frac{7}{2}z \right) 
\ln(z)-\frac{1}{2}(1+z) \ln^2(z)\right. \nonumber \\
& - & \left. 5(1-z) \right\}\nonumber \\
&+& C_F N_c\left\{ \left(\frac{1}{2}\ln^2(z)+\frac{11}{6}\ln(z)
\right)\delta P_{qq}(z)-\left( \frac{67}{18}-\frac{\pi^2}{6} \right)
(1+z)+(1+z)\ln(z) \right. \nonumber \\
&+& \left. \frac{20}{3}(1-z) \right\} 
+ \frac{C_F N_f}{2}\left\{ -\frac{2}{3}\ln(z)
\delta P_{qq}(z)+\frac{10}{9}(1+z)-\frac{4}{3}(1-z) \right\}\nonumber \\
& - & C_F \left( C_F-\frac{N_c}{2}\right) \left\{ 2
\delta P_{qq}(-z) S_2(z)+2 (1+z) \ln(z)+4(1-z) \right\},
\end{eqnarray}
\begin{eqnarray}
P_{qq}^-(z)&=& C_F^2\left\{ -\left( 2 \ln(z)\ln(1-z)+\frac{3}{2}\ln(z)
\right) \delta P_{qq}(z)-\left( \frac{3}{2}+\frac{7}{2}z \right)\ln(z)-
\frac{1}{2}(1+z)\ln^2(z)\right. \nonumber \\
&-& \left. 5(1-z) \right\} \nonumber \\
&+& C_F N_c \left\{ \left( \frac{1}{2}\ln^2(z)+\frac{11}{6}\ln(z) \right)
\delta P_{qq}(z)-\left( \frac{67}{18}-\frac{\pi^2}{6}\right)
(1+z)+(1+z)\ln(z) \right. \nonumber \\
&+& \left. \frac{20}{3}(1-z) \right\} 
+ \frac{C_F N_f}{2}\left\{ \frac{2}{3} \ln(z)
\delta P_{qq}(z)+\frac{10}{9}(1+z)-\frac{4}{3}(1-z) \right\} \nonumber \\
&+& C_F\left\{ C_F-\frac{N_c}{2}\right) \left\{ 2\delta P_{qq}(-z) S_2(z)+
2(1+z)\ln(z)+4(1-z) \right\},
\end{eqnarray}
\begin{eqnarray}
\delta P_{qq}(z)& =& \frac{2}{1-z} - 1 - z, \\
\delta P_{gg}(z)& =& \frac{1}{1-z} - 2 z + 1, \\
S_2 &= &-2 {\rm Li_2}(-z)-2 \ln(z) \ln(1+z)+\frac{1}{2}\ln(z)^2-\frac{\pi^2}{6}
\end{eqnarray}
with
\begin{equation}
{\rm Li_2}(-z)\equiv - \int_0^{-z} {{dx}\over x}\ln(1-x)=\sum_{n=1}^{\infty} 
{{(-z)^n}\over {n^2}},
\end{equation}
\begin{eqnarray}
P^{(1)}_{gq}(z)&=& \frac{C_F N_f}{2}\left\{ 
-\frac{4}{9}(z+4)-\frac{4}{3}(2-z)\ln(1-z)\right \} 
+ C_F^2\left\{ -\frac{1}{2}-\frac{1}{2}(4-z)\ln(z)\right. \nonumber \\
&-& \left. (2+z)\ln(1-z)+
\left( -4-\ln^2(1-z)+\frac{1}{2}\ln^2(z)\right)(2-z) \right\} \nonumber \\
&+ & C_F N_c\left\{ 
(4-13z)\ln(z)+(10+z)\frac{\ln(1-z)}{3}+\frac{(41+35z)}{9}+\frac{1}{2}
\left( -2 S_2(z) \right. \right. \nonumber \\
&+& \left. \left. 3\ln^2(z)\right) 
\times  (2+z)+\left( \ln^2(1-z)-2\ln(1-z)\ln(z)-
\frac{\pi^2}{6}\right) (2-z) \right\},
\end{eqnarray}
\begin{eqnarray}      
P_{qg}^{(1)}(z)&=& \frac{C_F N_f}{2}\left\{ -22+27z-9\ln(z)+8(1-z)\ln(1-z)+
(2z-1) \right. \nonumber \\
& & \left. (2\ln^2(1-z)-4\ln(1-z)\ln(z) 
+ \ln^2(z)-\frac{2\pi^2}{3} \right\}
\nonumber \\
&+& \frac{N_c N_f}{2}\left\{ 2(12-11z)-8(1-z)\ln(1-z)
+2(1+8z)\ln(z)-2\left( \ln^2(1-z)-\frac{\pi^2}{6}\right)\right.
\nonumber \\
&\times & \left. (2z-1)-
(2S_2(z)-3\ln^2(z))(-2z-1) \right\},
\end{eqnarray}
and
\begin{eqnarray}
P_{gg}^{'(1)}(z)&=&-\frac{N_c N_f}{2}\left\{ 4(1-z)+\frac{4}{3}(1+z)\ln(z)+
\frac{20}{9}(1-2z) \right\} \nonumber \\
&-& \frac{C_F N_f}{2}\left\{  10(1-z)+2(5-z)\ln(z)+2(1+z)
\ln^2(z)\right\} \nonumber \\
& +& N_c^2\left\{ (29-67z)\frac{\ln(z)}{3}-\frac{19}{2}(1-z)+4
(1+z)\ln^2(z)-2S_2(z)\delta P_{gg}(-z) \right. \nonumber \\
&+& \left. \left( \ln^2(z)-4\ln(1-z)\ln
(z)\right) \delta P_{gg}(z)+\left( \frac{67}{9}-\frac{\pi^2}{3}
\right)(1-2z) \right\}.
\end{eqnarray}     
\end{appendix}

\pagebreak

\pagebreak

\noindent
\begin{center}
{\large Figure Captions}
\end{center}
\newcounter{num}
\begin{list}%
{[\arabic{num}]}{\usecounter{num}
    \setlength{\rightmargin}{\leftmargin}}

\item (a) Polarized gluon distribution $x \Delta G(x,Q^2)$ at $Q^2=100$
GeV$^2$ showing the GSA, GGRA, GGRB and GGRC parametrizations. (b) same
as (a) for the $\Delta \overline{u}$-quark distribution. (c) Same as (a) and
(b) for the strange quark distribution. The dotted curve near the solid
(GGRA) is for the LO version of GGRA. Same as (c) for the valence
u-quark distribution. Again the dotted curve near the solid curve is the
LO version of GGRA. The valence distributions for GGRA, GGRB and GGRC
cannot be distinguished.
\item (a) $p_T$-distribution for unpolarized prompt photon production at
$y^{\gamma}=0$ for $\sqrt{S}=200$ GeV. (b) Asymmetry $A_{LL}^{\gamma}$ for the
cross section in (a) as predicted by the GSA, GGRA, GGRB and GGRC
polarized parton distributions. (c) Asymmetries from (b) for GSA, GGRA
amd GGRC showing the contributions from the $qg$ and $q\overline{q}$
initial state contributions.
\item (a) Rapidity distribution for unpolarized prompt photon production
at $p_T^{\gamma} = 15$ GeV and $\sqrt{S}=200$ GeV. (b) Asymmetries for the
rapidity distribution as predicted by various polarized parton
parametrizations. 
\item (a) $p_T$-distribution for unpolarized prompt photon production at
$y^{\gamma}=0$ for $\sqrt{S}=40$ GeV. (b) Asymmetry $A_{LL}^{\gamma}$ for the
cross section in (a) as predicted by the GSA, GGRA, GGRB and GGRC
polarized parton distributions. 
\item (a) $p_T$-distribution for the unpolarized jet cross section at
$y^{J}=0$ for $\sqrt{S}=200$ GeV. (b) Asymmetry $A_{LL}^J$ for the total
and for various initial state contributions as predicted by the GSA
distributions. (c), (d) and (e): Same as (b) for the GGRA, GGRB and GGRC
parametrizations respectively. 
\item (a) $p_T$-distribution for the unpolarized jet cross section at
$y^{J}=0$ for $\sqrt{S}=500$ GeV. (b) Asymmetry $A_{LL}^J$ for the cross
section as  predicted by the GSA, GGRA, GGRB and GGRC polarized distributions. 
\end{list}

\pagebreak 
\epsffile{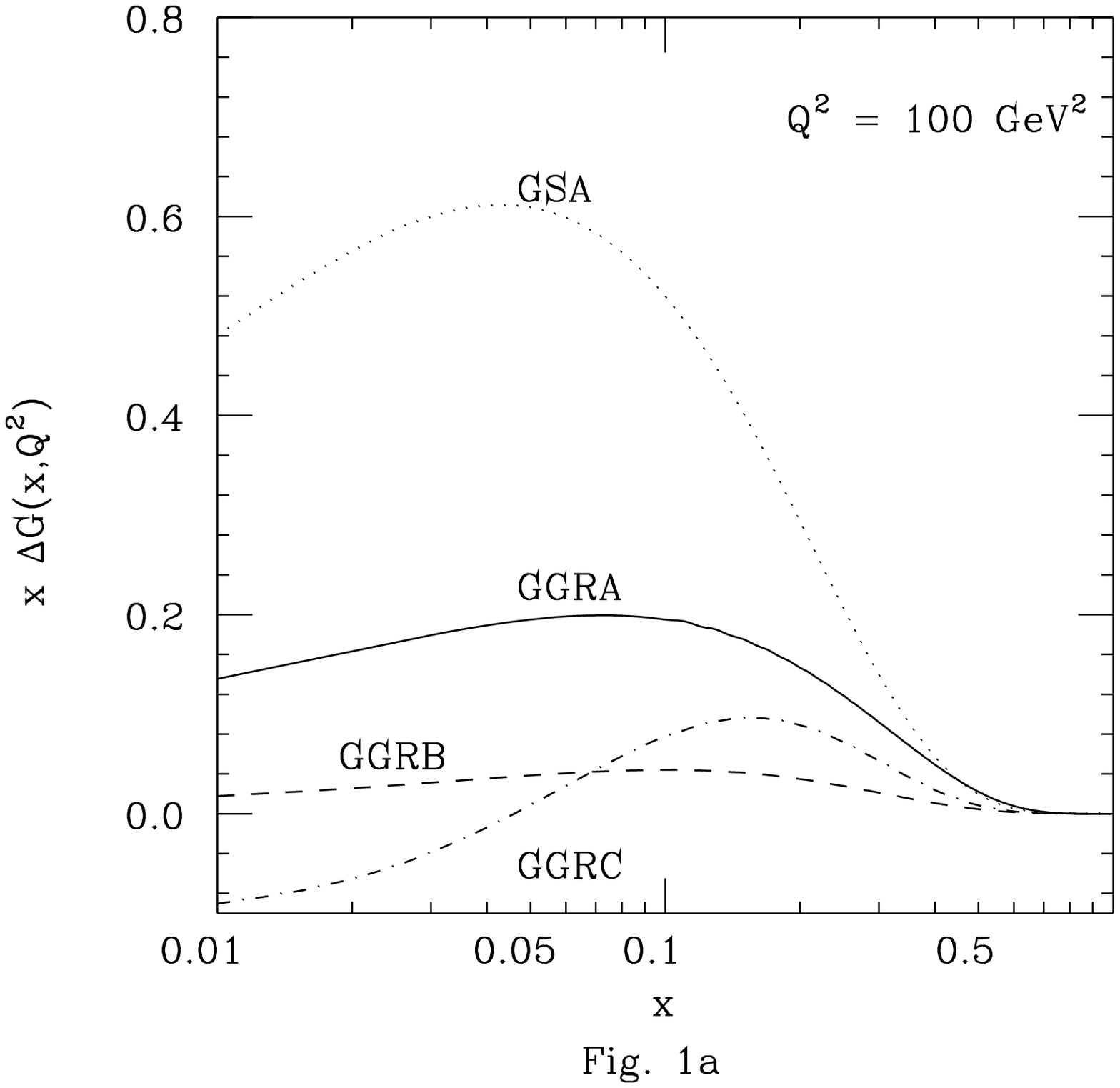}
\epsffile{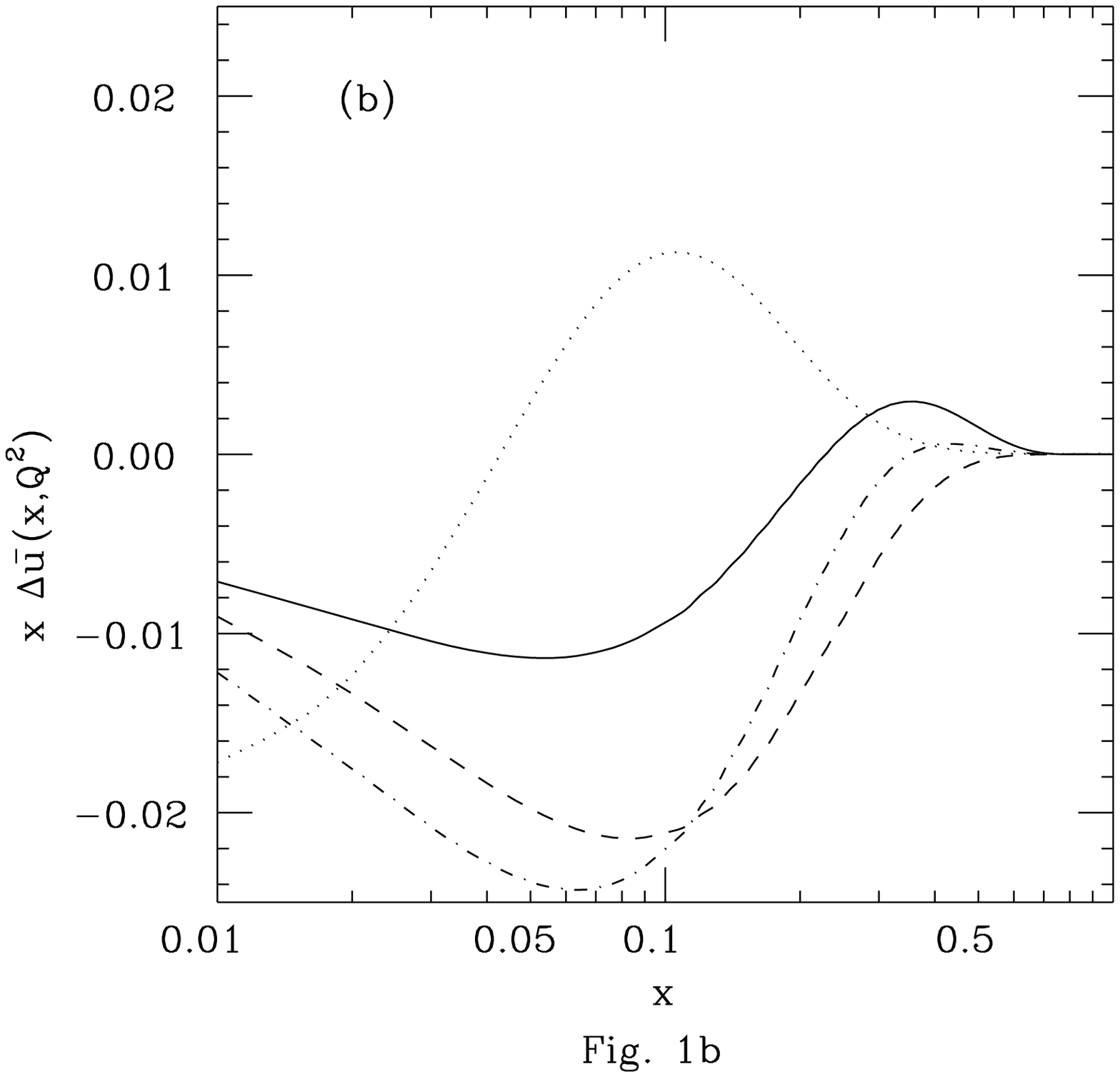} 
\epsffile{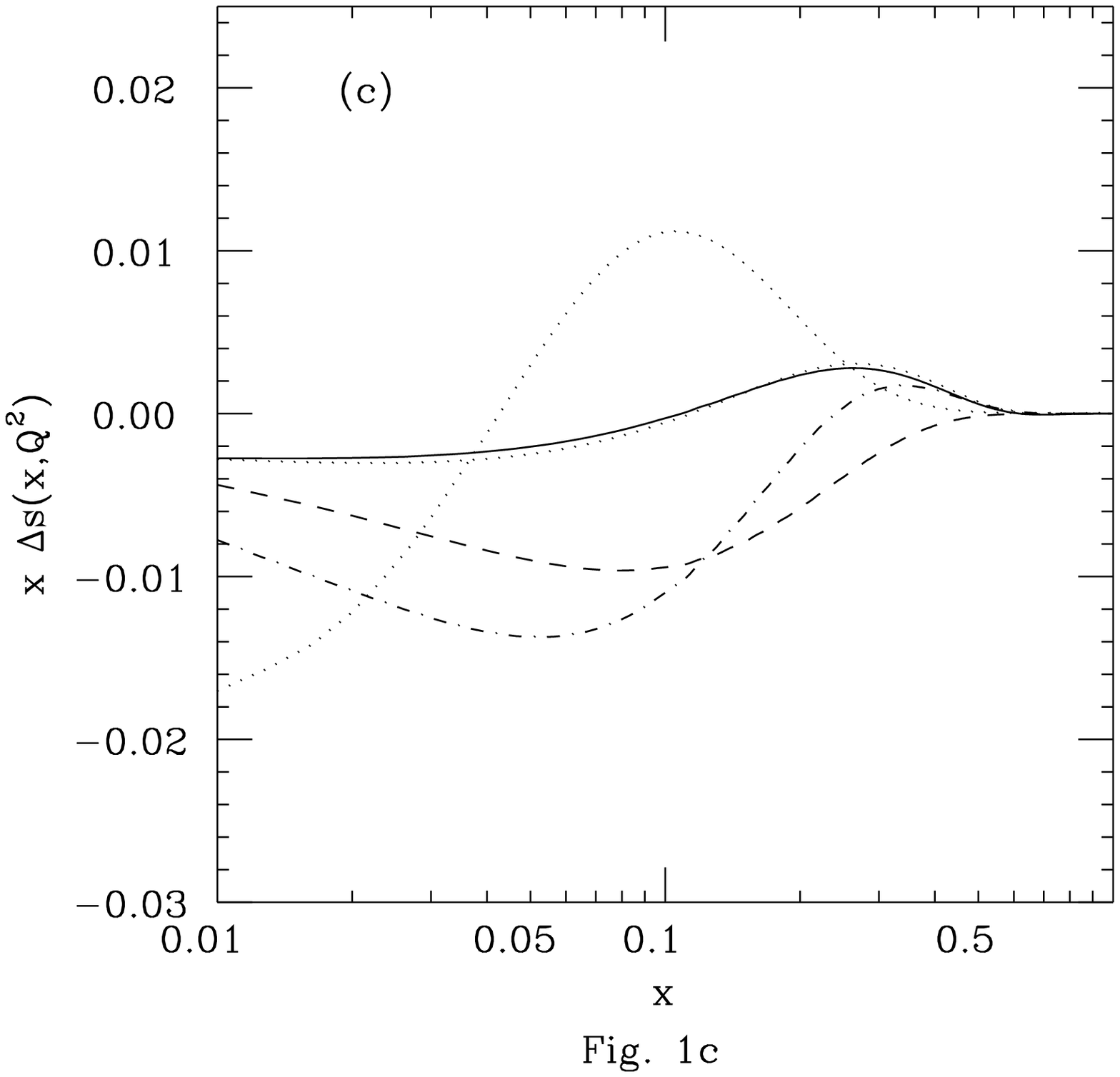} 
\epsffile{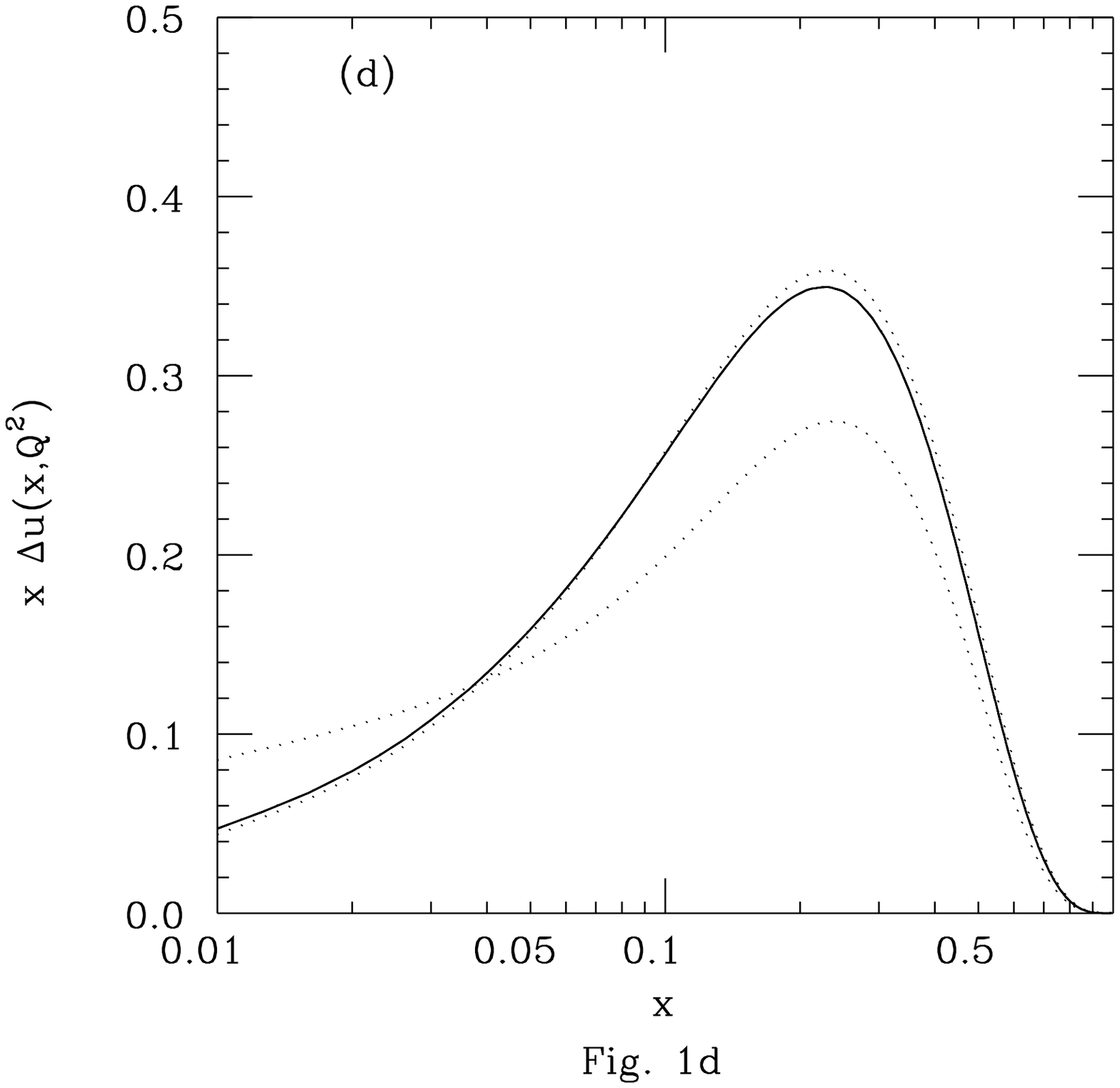}
\epsffile{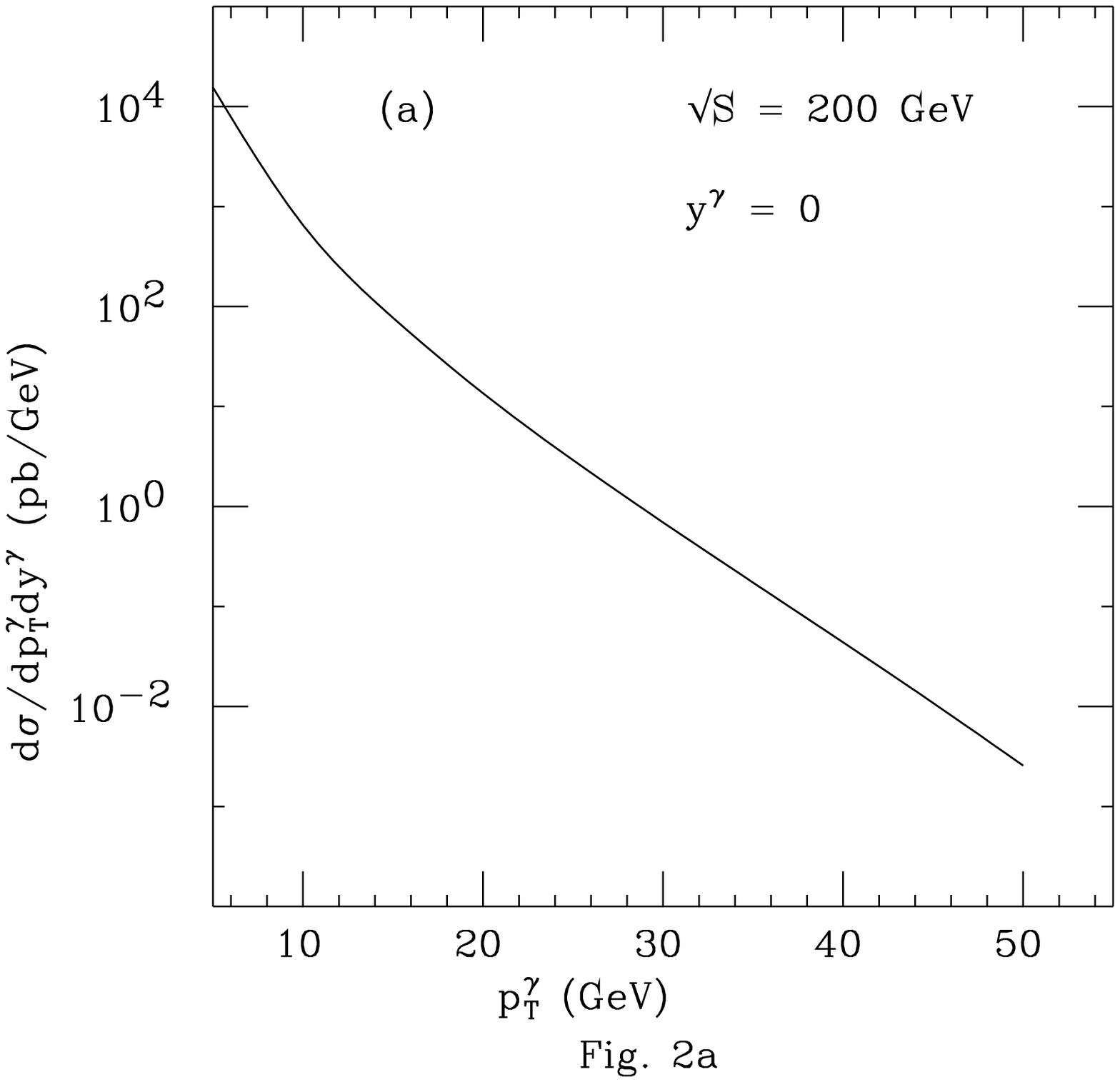} 
\epsffile{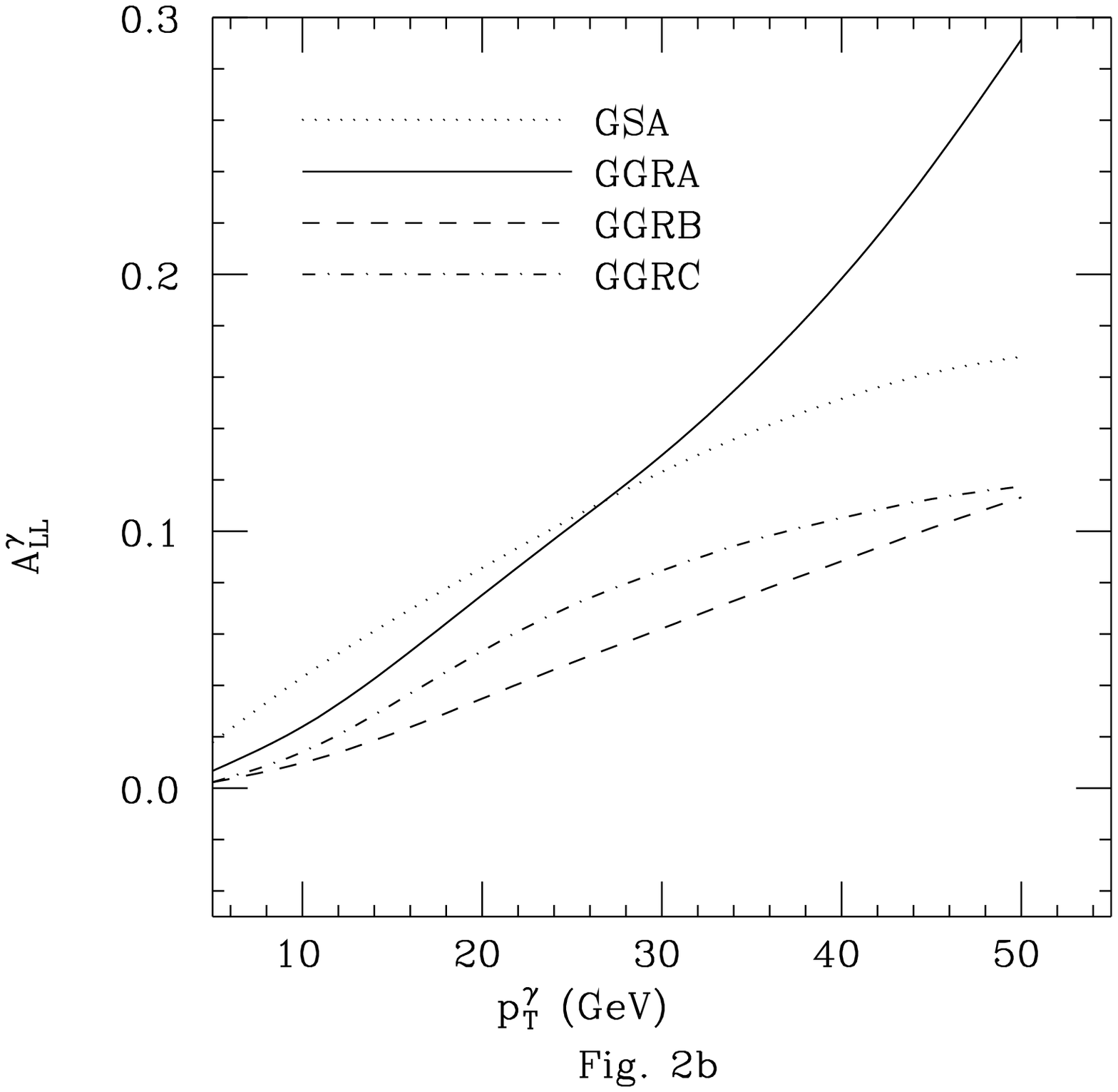} 
\epsffile{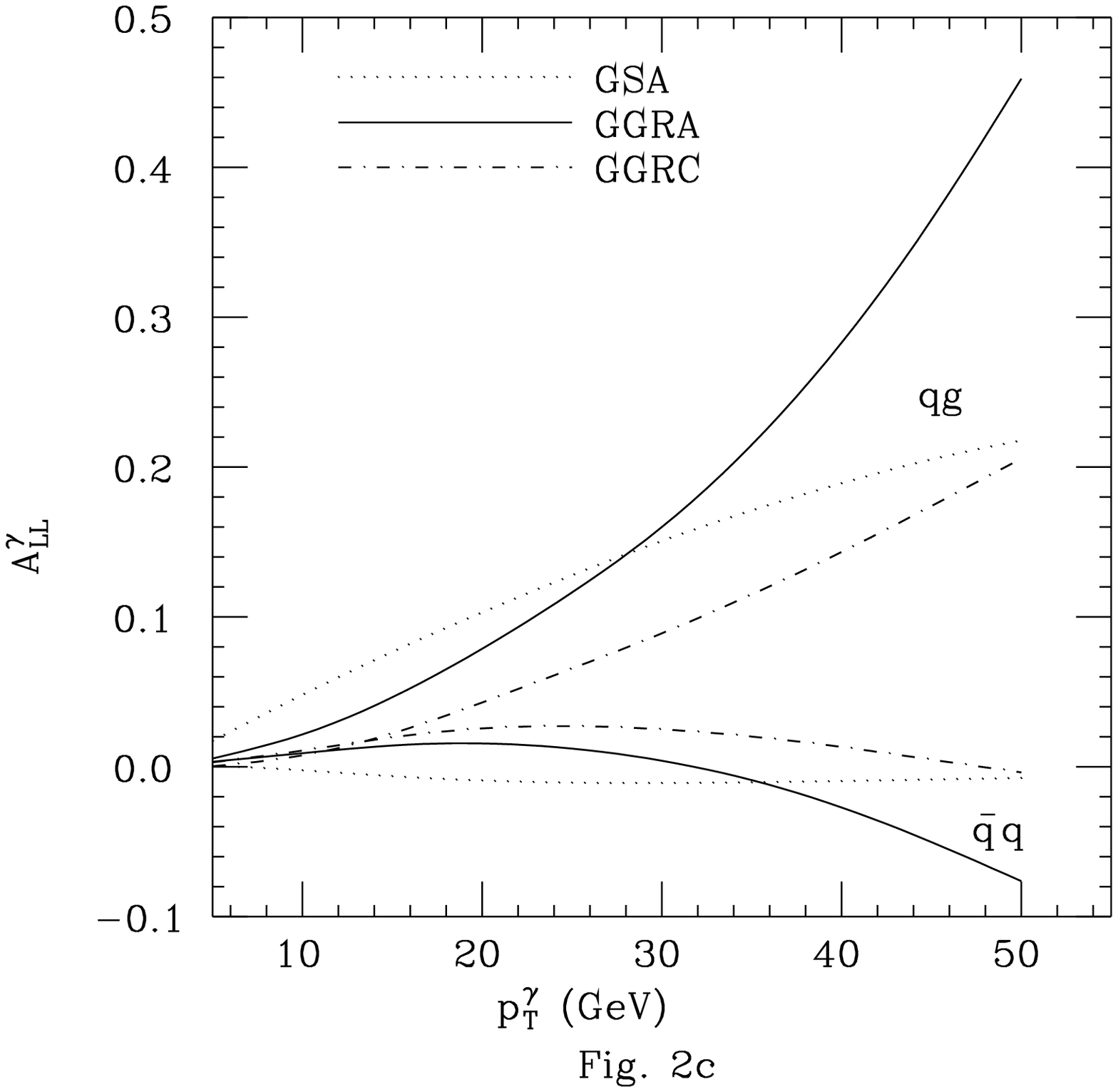} 
\epsffile{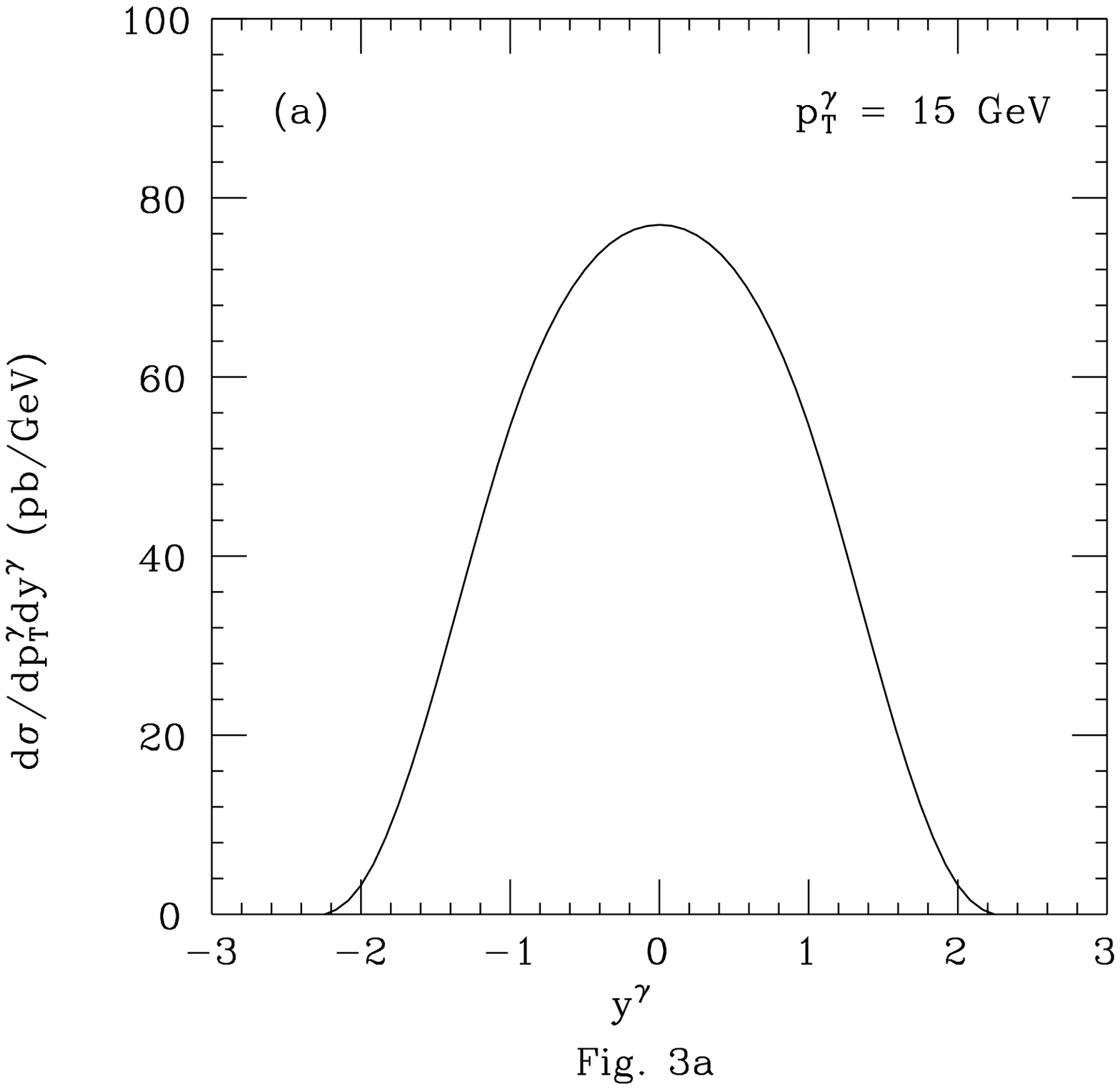} 
\epsffile{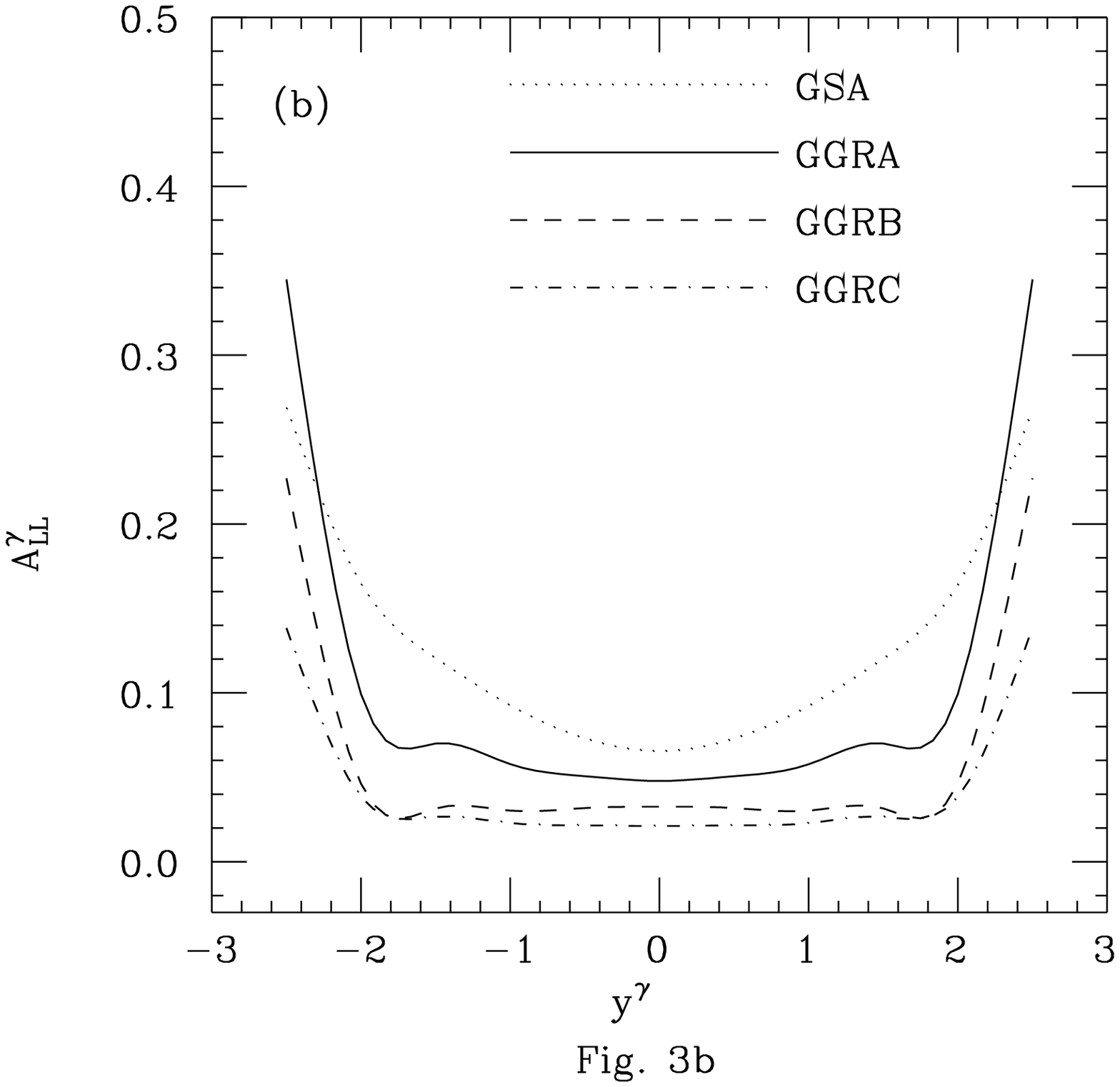} 
\epsffile{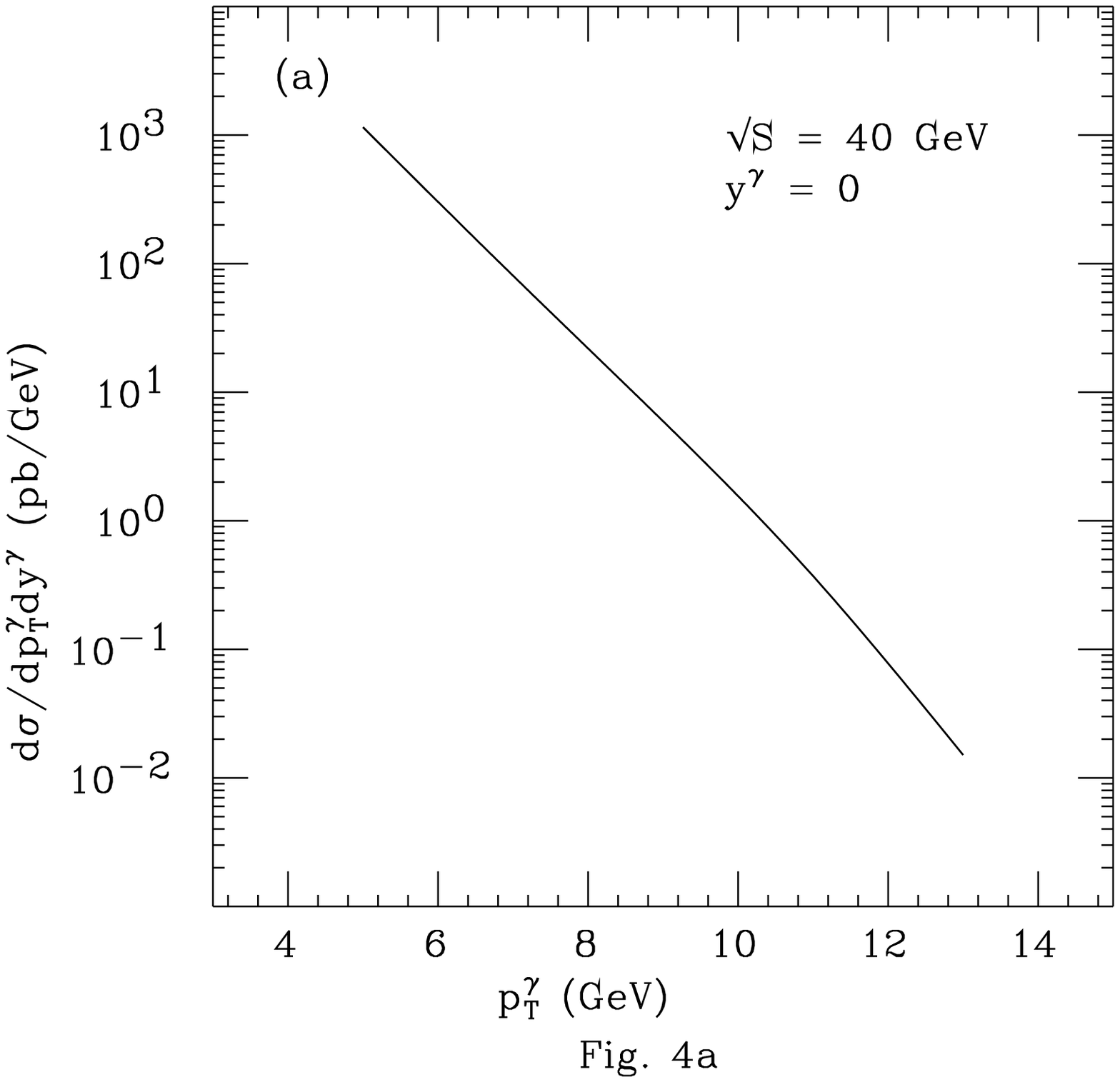}
\epsffile{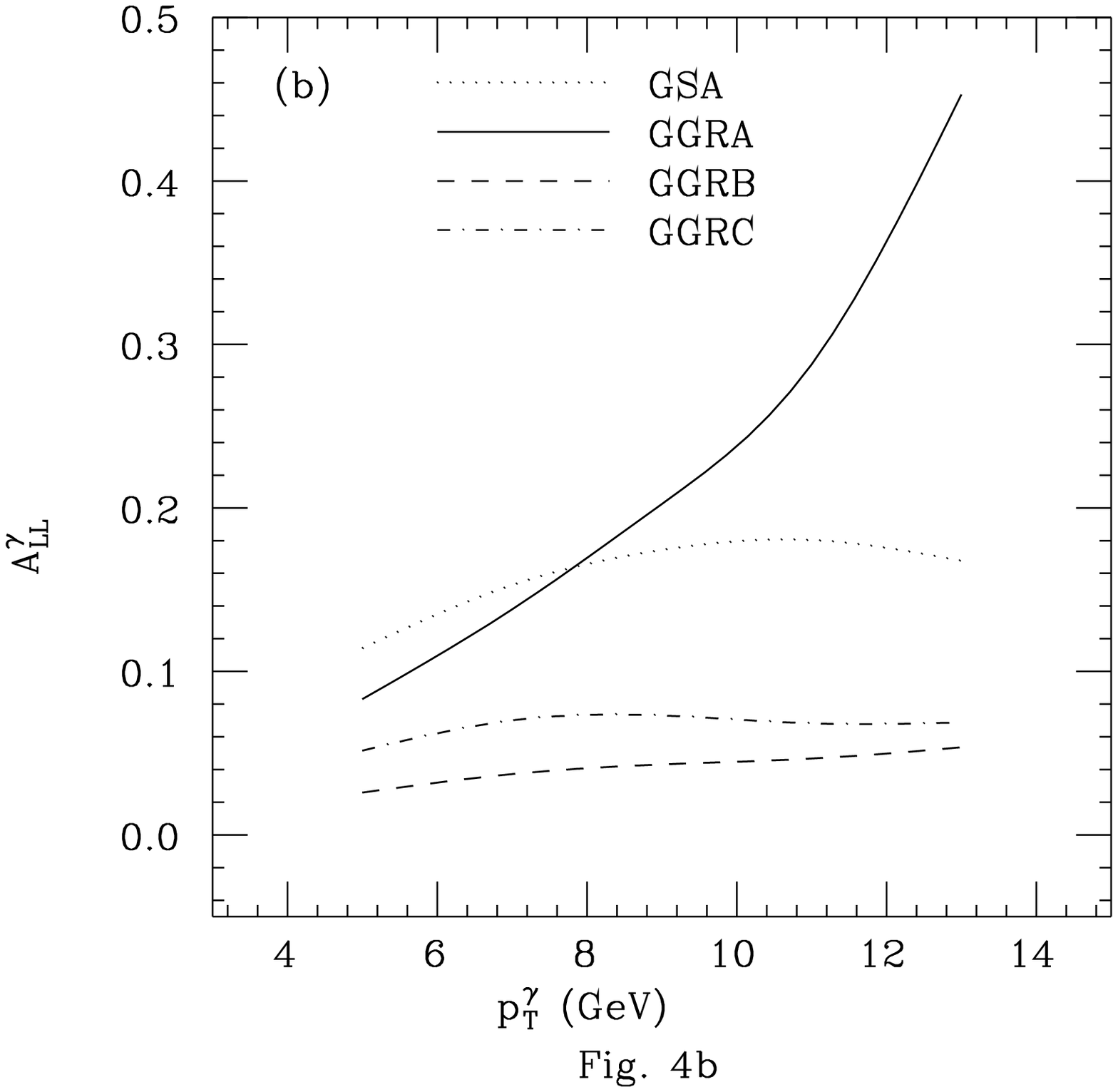} 
\epsffile{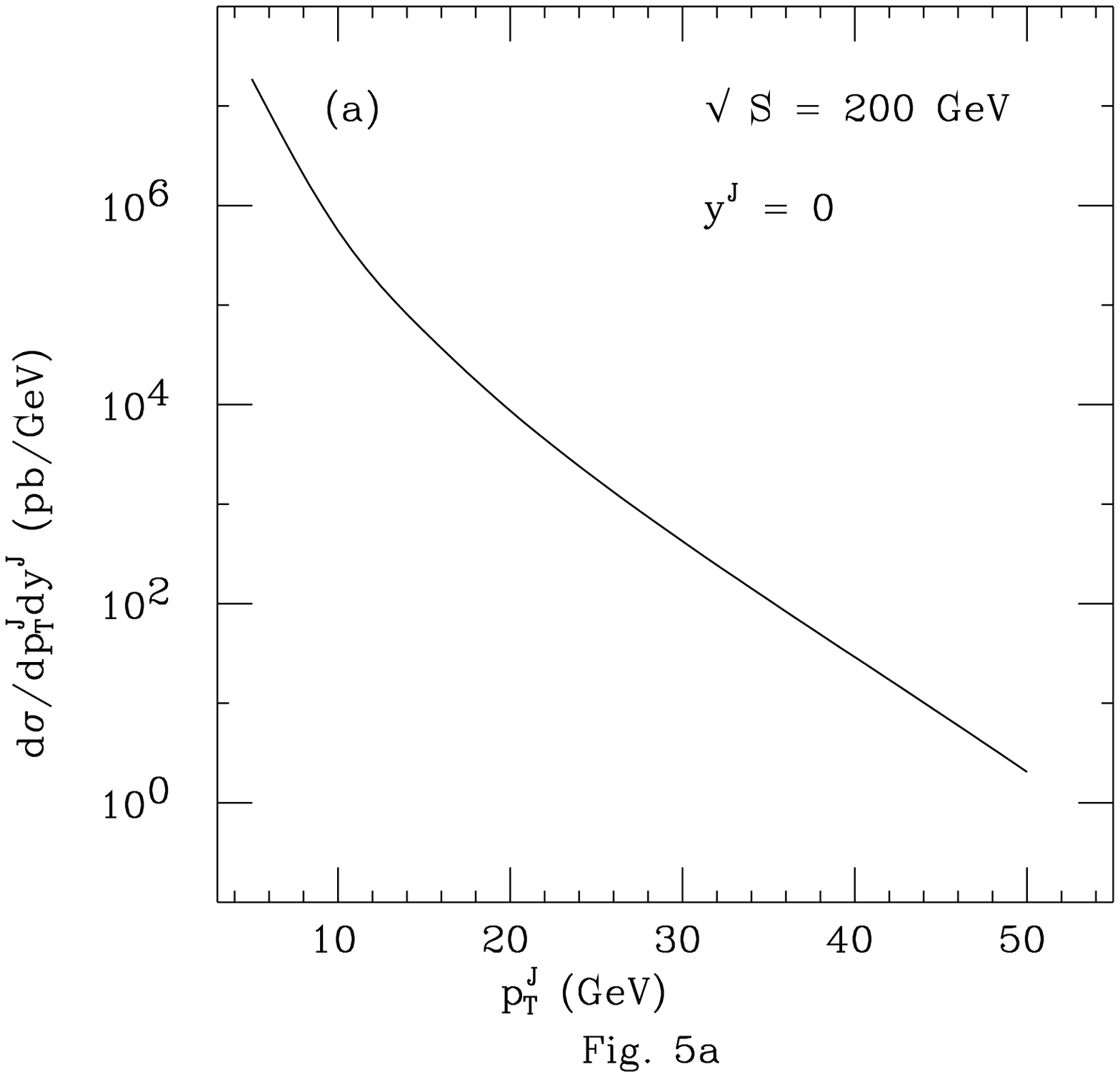} 
\epsffile{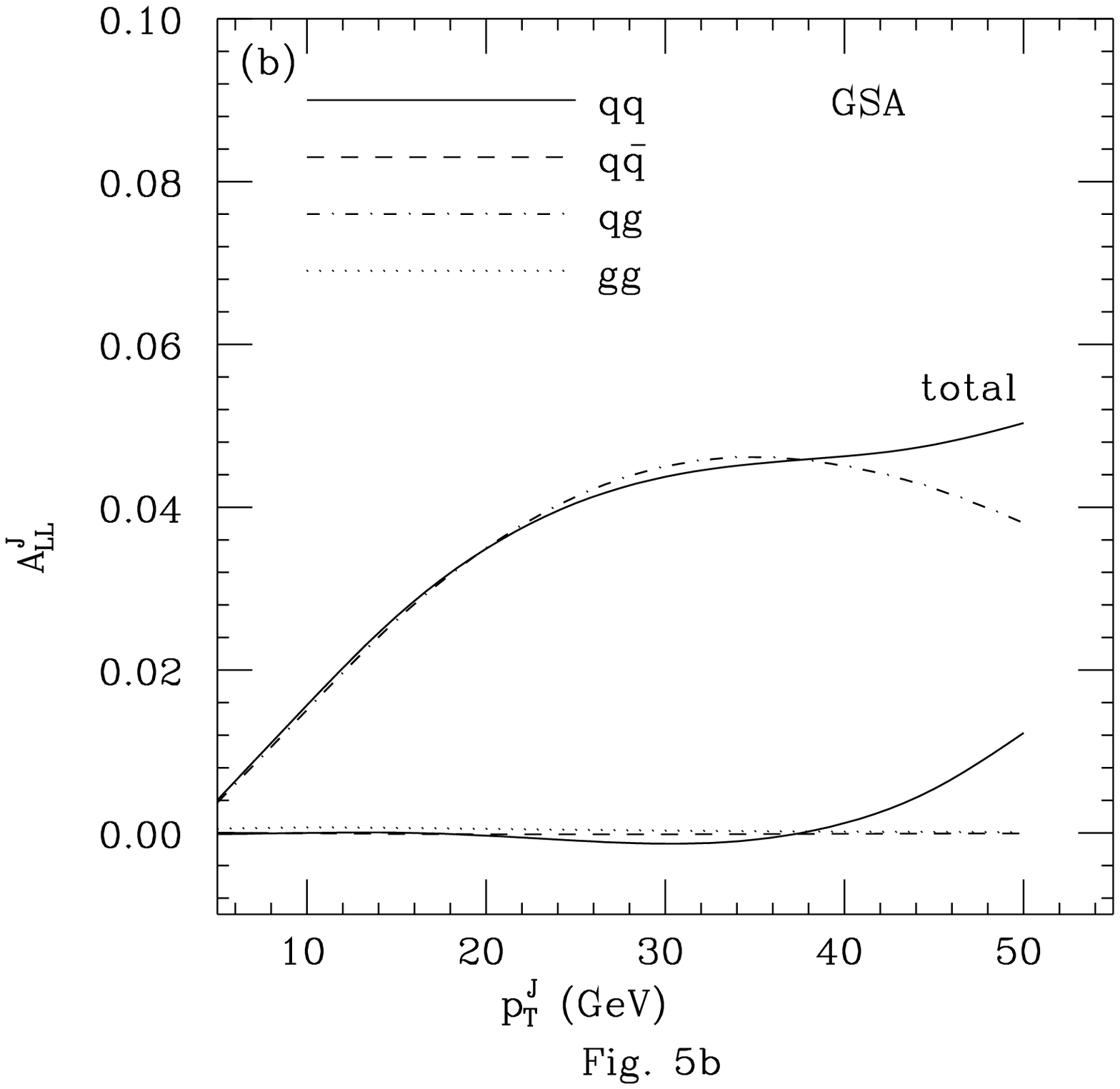}
\epsffile{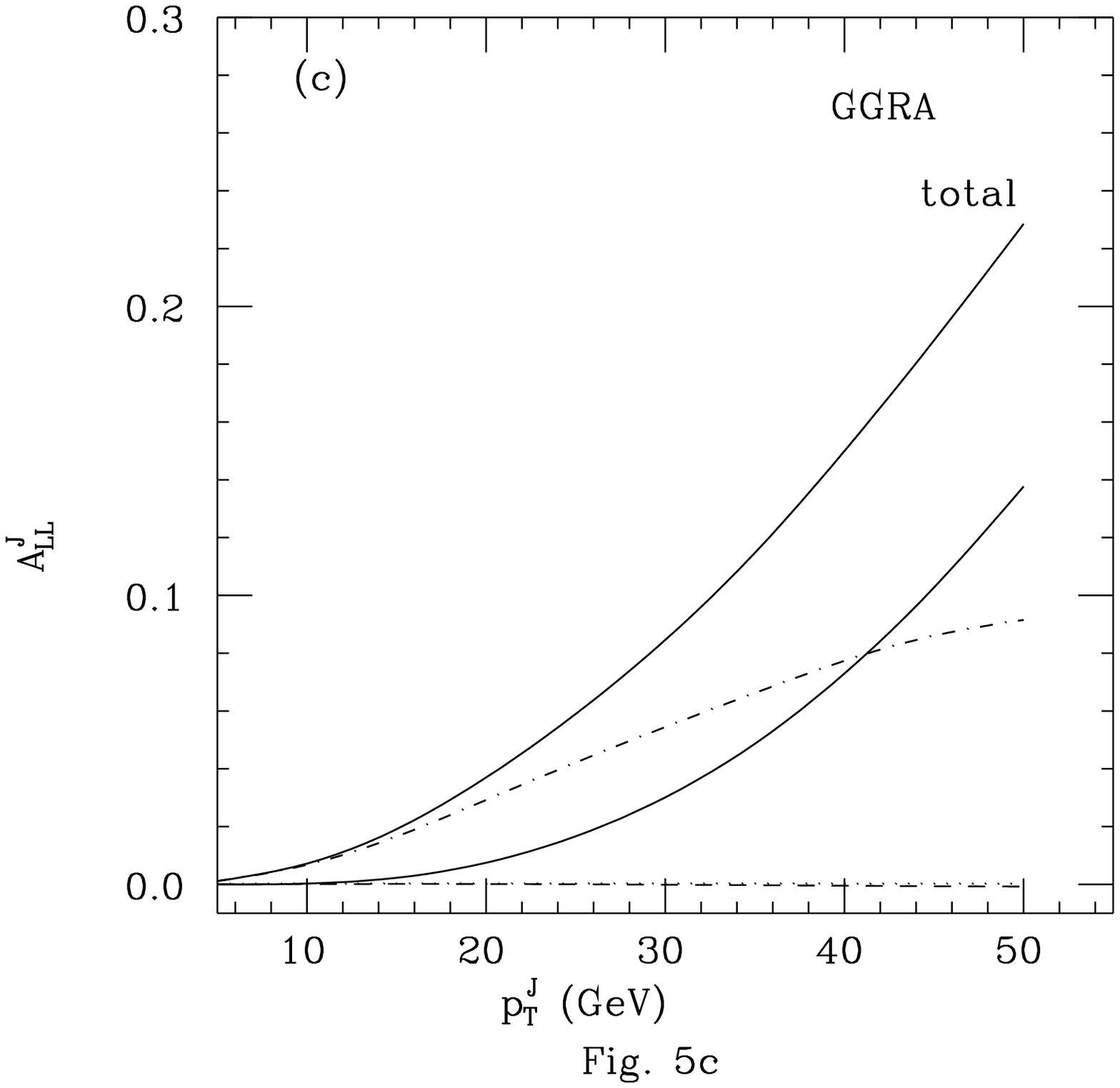} 
\epsffile{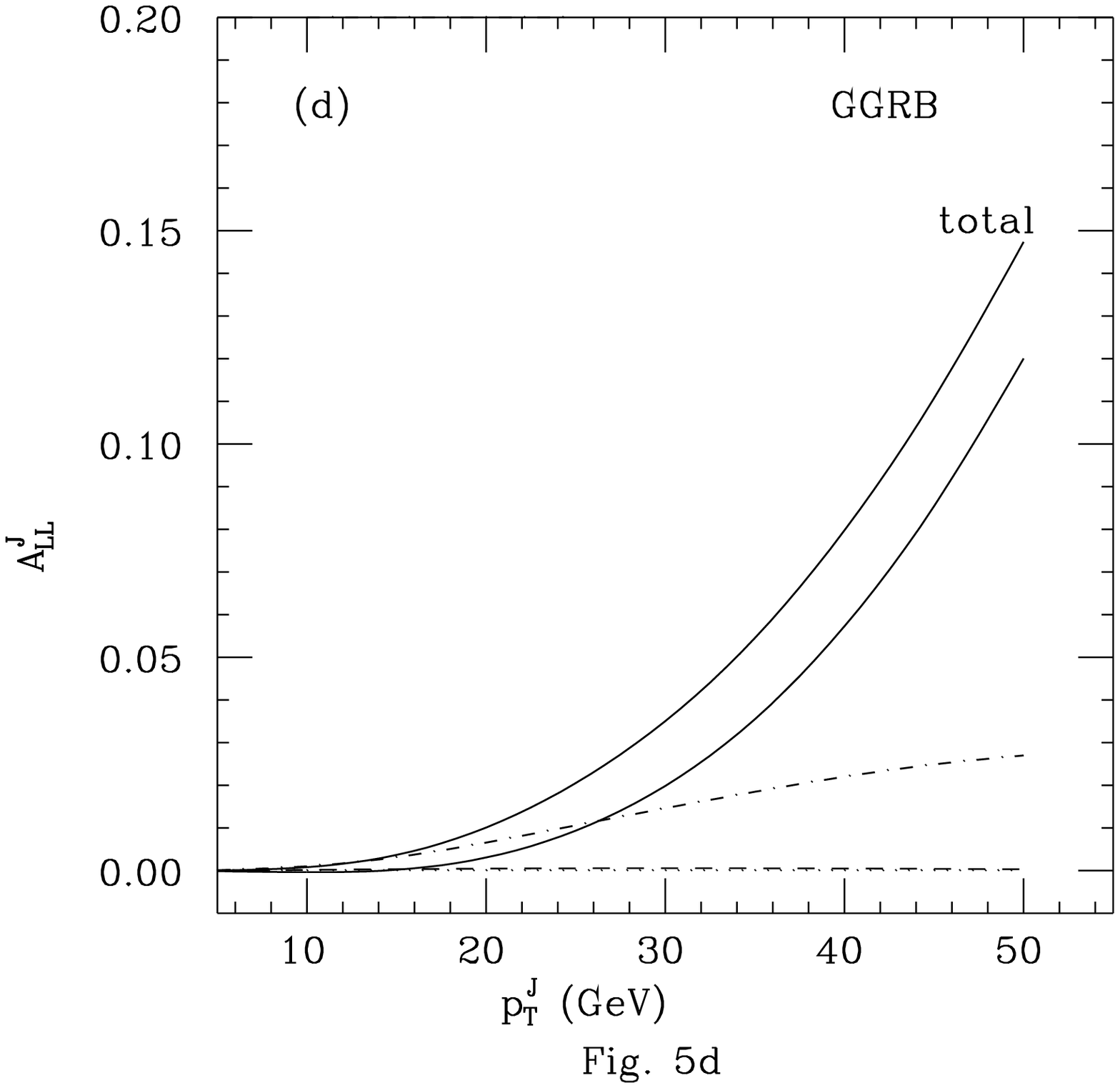} 
\epsffile{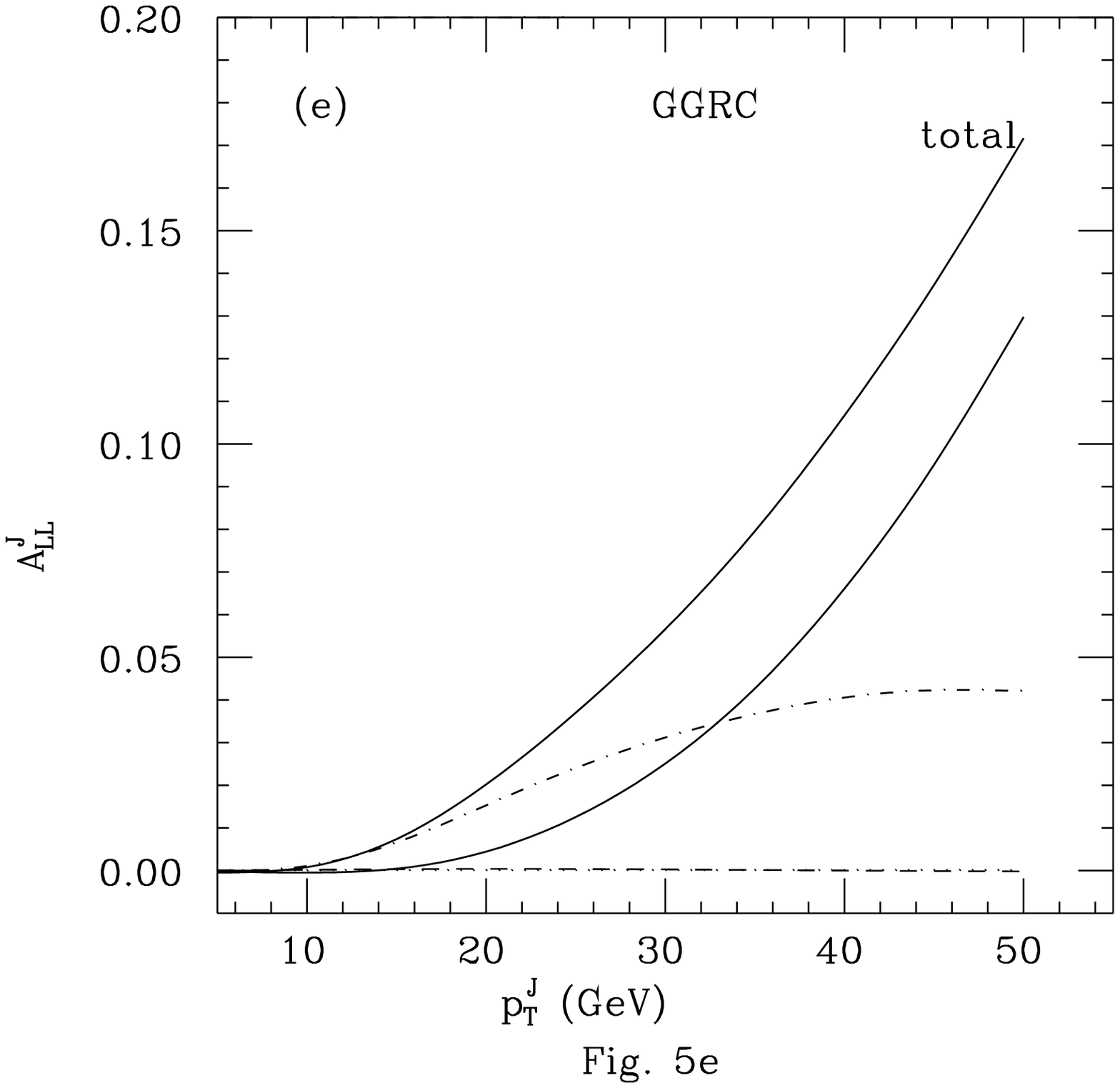} 
\epsffile{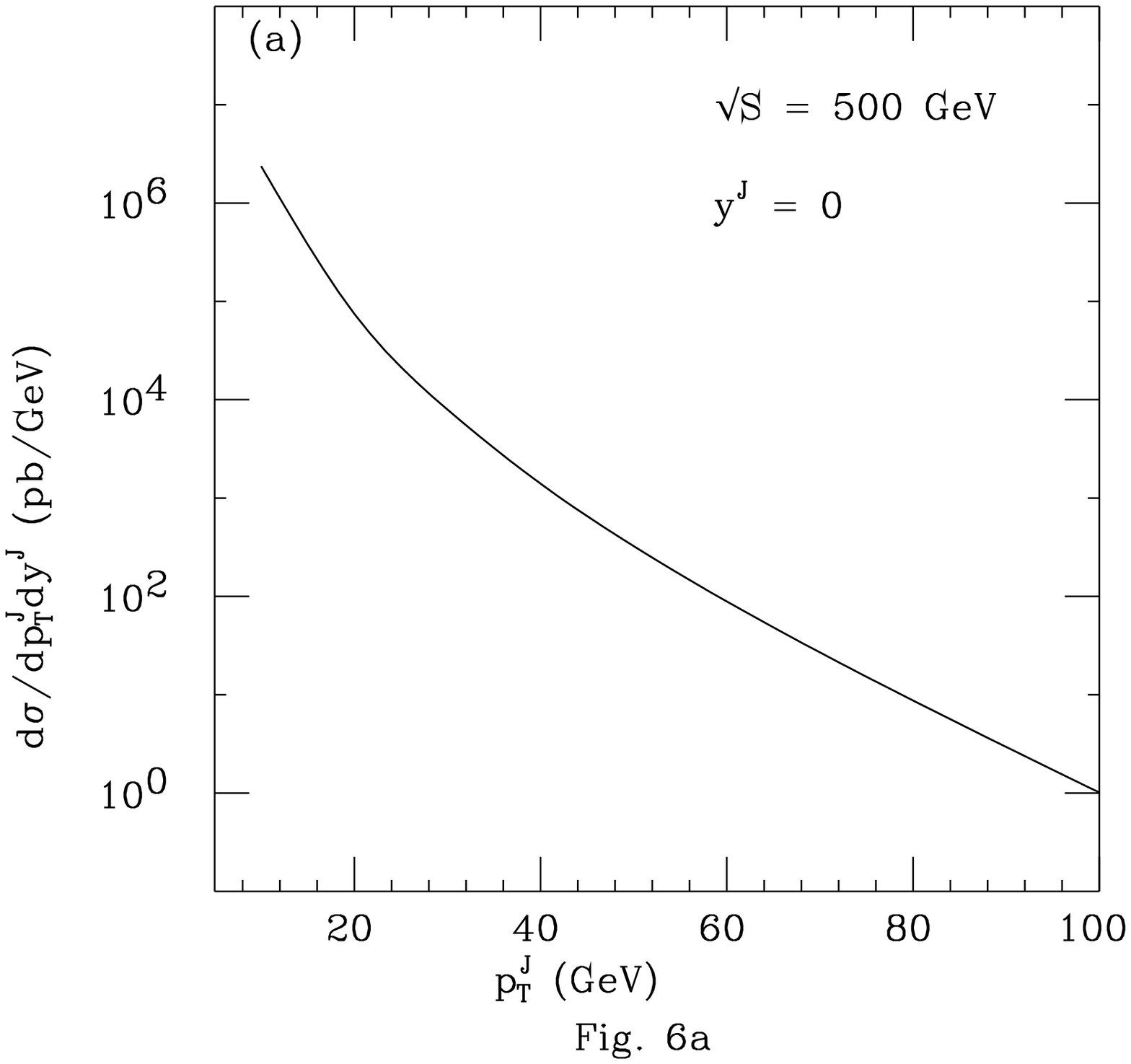} 
\epsffile{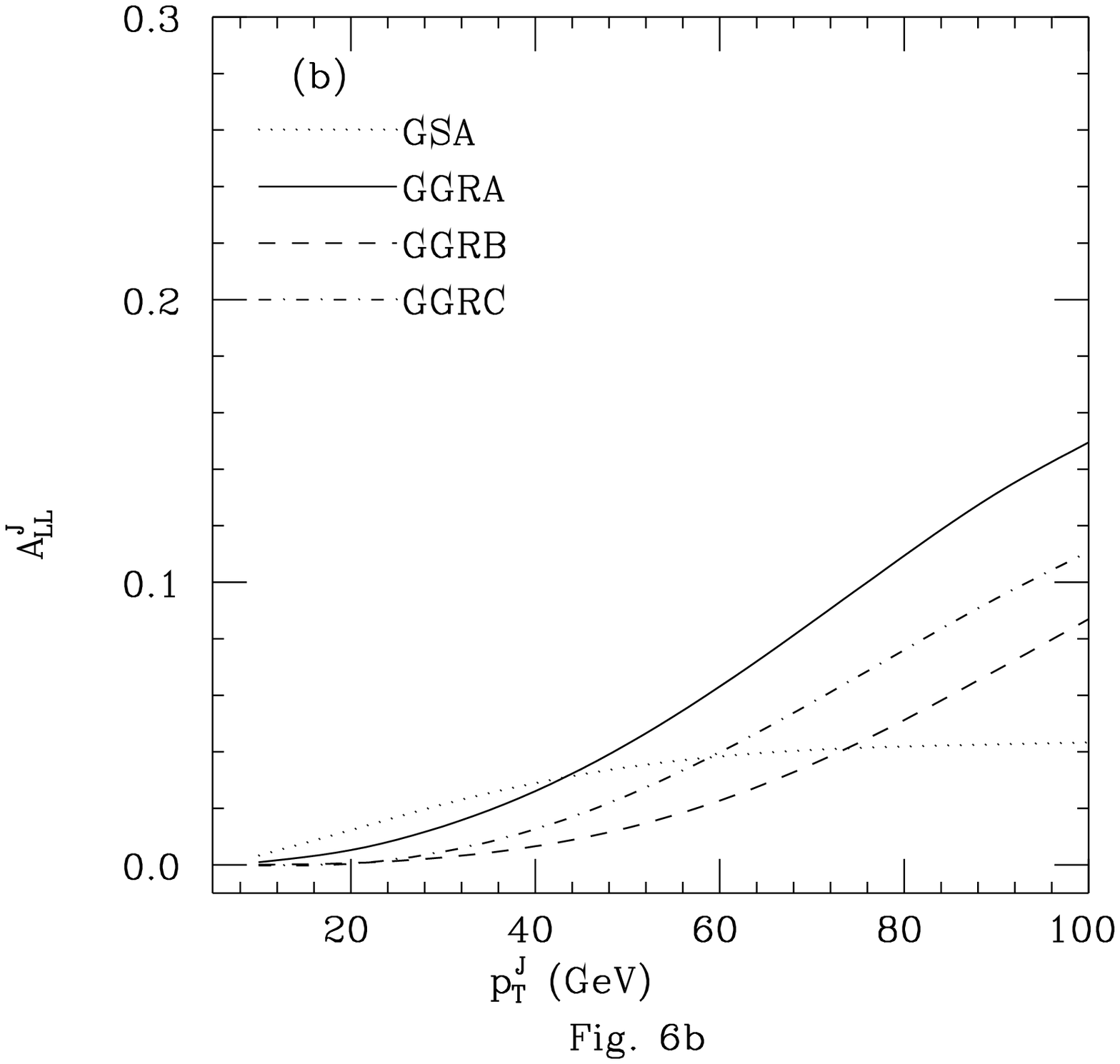} 

\begin{references}
\bibitem{gr} M. Goshtasbpour and G. P. Ramsey, Phys. Rev. {\bf D55}, 1244
(1997).
\bibitem{ggr} L. E. Gordon, M. Goshtasbpour and G. P. Ramsey, Phys. Rev. {\bf
D58}, 094017 (1998), (hep-ph/9803351).
\bibitem{CTEQ} CTEQ Collaboration, H. L. Lai {\it et al.,} Phys. Rev.
{\bf D51}, 4763 (1995). 
\bibitem{werner} R. Mertig and W. L. van Neerven, Z. Phys. {\bf C70},
637 (1996); W. Vogelsang, Phys. Rev. {\bf D54}, 2023 (1996) and Nucl. Phys.
{\bf B475}, 47 (1996).
\bibitem{rossi} G. Rossi, Phys. Rev. {\bf D29}, 852 (1984).
\bibitem{bnl} Proceedings from the RIKEN BNL Research Center Workshop, BNL,
April 27-29, 1998, G. Bunce, {\it et. al.,} Editors; BNL-65615, vol. 7.
See also Web site- http://www.rhic.bnl.gov for more information.
\bibitem{gluon} R. D. Ball, S. Forte and G. Rudolfi, Nucl. Phys. {\bf B496},
337 (1997); see also Altarelli, {\it et. al.,} hep-ph/9707276 and
Acta Phys. Polon. {\bf B29}, 1145 (1998), (hep-ph/9803237); and R. L. Jaffe,
Phys. Lett. {\bf B365}, 359 (1996).
\bibitem{nowak} W.-D. Nowak "Possible Measurements of Single and Double
Spin Asymmetries with HERA-$\vec{\rm N}$", hep-ph/9605411, published in the
proceedings of the Trieste Collider Spin Physics Workshop, 169 (1995).
Also see Nucl. Phys. {\bf A622}, 78c (1997).
\bibitem{bergerqiu} E. L. Berger and J. Qiu, Phys. Rev. {\bf D40}, 778
(1989).
\bibitem{contogouris} A. P. Contogouris, B. Kamal, Z. Merebashvili and F. V.
Tkachov, Phys. Lett. {\bf B304}, 329 (1993); Phys. Rev. {\bf D48}, 4092
(1993). 
\bibitem{gorvogel} L. E. Gordon and W. Vogelsang, Phys. Rev. {\bf D48},
3136 (1993) and {\bf D49}, 170 (1994).
\bibitem{gordon1} L. E. Gordon, Nucl. Phys. {\bf B501}, 175 (1997) and
{\it ibid}., p. 197.
\bibitem{gordon2} L. E. Gordon, Phys. Lett. {\bf B406}, 184 (1997).
\bibitem{nlojet} D. de Florian, S. Frixione, A. Signer and W. Vogelsang,
hep-ph/9808262.
\bibitem{gs} T. Gehrmann and W. J. Stirling, Phys. Rev. {\bf D53}, 6100
(1996).
\bibitem{goto} Y. Goto, see article in ref \cite{bnl}; \\ 
transparencies available
at URL http://www.phenix.bnl.gov/WWW/publish/goto/.
\bibitem{soffer} J. M. Virey and J. Soffer, Nucl. Phys. {\bf B509}, 297 (1998)
and C. Bourrely and J. Soffer, Nucl. Phys. {\bf B445}, 341 (1995).
\bibitem{st} G. P. Skoro and M. V. Tokarev, Nuovo Cim. {\bf 111A}, 353 (1998).
\bibitem{tt} O. Teryaev and A. Tkabladze, Phys. Rev. {\bf D56}, 7331 (1997).
\bibitem{gv2} L. E. Gordon and W. Vogelsang, Phys. Lett. {\bf B387}, 629
(1996).
\bibitem{ddhlr} A. De Roeck, A. Deshpande, V. W. Hughes, J. Lichtenstadt,
and G. Radel, hep-ph/9801300, DESY 97-249, to appear in Eur. Phys. J. {\bf C}.
\end{references}
\end{document}